\documentclass[fleqn,usenatbib]{mnras}

\usepackage{newtxtext,newtxmath}

\usepackage[T1]{fontenc}
\usepackage{ae,aecompl}
\usepackage{lscape}

\usepackage{graphicx}	
\usepackage{amsmath}	

\newcommand{\NGTS}{NGTS}

\newcommand{\kms}{km\,s$^{-1}$}
\newcommand{\ms}{m\,s$^{-1}$}
\newcommand{\masy}{mas\,y$^{-1}$}
\newcommand{\mpl}{\mbox{M$_{p}$}}
\newcommand{\rpl}{\mbox{R$_{p}$}}
\newcommand{\mstar}{\mbox{M$_{s}$}}
\newcommand{\rstar}{\mbox{R$_{s}$}}
\newcommand{\mjup}{\mbox{M$_{\rm J}$}}
\newcommand{\rjup}{\mbox{R$_{\rm J}$}}
\newcommand{\msun}{\mbox{M$_{\odot}$}}
\newcommand{\rsun}{\mbox{R$_{\odot}$}}

\newcommand{\gccc}{g\,cm$^{-3}$}

\newcommand{\Nstar}{NGTS-21}

\newcommand{\Nplanet}{NGTS-21b}

\title[\Nplanet]{\Nplanet: An Inflated Super-Jupiter Orbiting a Metal-poor K dwarf}

\author[Alves, D. et al.]{
\parbox{\textwidth}{
Douglas R. Alves,$^{1,3}$\thanks{E-mail: \href{douglasalvesastro12@gmail.com}{douglasalvesastro12@gmail.com}}
James S. Jenkins,$^{2,3}$
Jose I. Vines,$^{1}$
Louise D. Nielsen,$^{4}$
Samuel Gill,$^{5,6}$
Jack S. Acton,$^{8}$
D.~R.~Anderson$^{5,6}$
Daniel~Bayliss,$^{5,6}$
Fran\c{c}ois Bouchy,$^{11}$
Hannes~Breytenbach,$^{9,10}$
Edward~M. Bryant,$^{16}$
Matthew R. Burleigh,$^{8}$
Sarah L. Casewell,$^{8}$
Philipp~Eigm\"uller,$^{7}$
Edward~Gillen,$^{13,14}$
Michael R.~Goad,$^{8}$
Maximilian~N.~G{\"u}nther,$^{17}$
Beth A. Henderson,$^{8}$
Alicia Kendall,$^{8}$
Monika Lendl,$^{11}$
Maximiliano~Moyano,$^{12}$
Ramotholo~R.~Sefako,$^{10}$
Alexis~M.~S.~Smith,$^{7}$
Jean C. Costes,$^{18}$
Rosanne H. Tilbrook,$^{8}$
Jessymol~K.~Thomas,$^{10}$
St\'{e}phane~Udry,$^{11}$
Christopher~A.~Watson,$^{15}$
Richard~G.~West,$^{5,6}$
Peter~J.~Wheatley,$^{5,6}$
Hannah~L.~Worters,$^{10}$
Ares Osborn$^{5,6}$
}
\vspace{3mm}
\\
$^1$Departamento de Astronom\'ia, Universidad de Chile, Casilla 36-D, Santiago, Chile\\
$^2$N\'ucleo de Astronom\'ia, Facultad de Ingenier\'ia y Ciencias, Universidad Diego Portales, Av. Ej\'ercito 441, Santiago, Chile\\
$^3$Centro de Astrof\'isica y Tecnolog\'ias Afines (CATA), Casilla 36-D, Santiago, Chile\\
$^{4}$European Southern Observatory (ESO), Karl-Schwarzschild-Stra{\ss}e 2, 85748 Garching bei M{\"u}nchen, Germany\\
$^{5}$ Department of Physics, University of Warwick, Gibbet Hill Road, Coventry CV4 7AL, UK \\
$^{6}$ Centre for Exoplanets and Habitability, University of Warwick, Gibbet Hill Road, Coventry CV4 7AL, UK\\
$^{7}$Department of Extrasolar Planets and Atmospheres, Institute of Planetary Research, German Aerospace Center (DLR), Rutherfordstra\ss e 2, 12489 Berlin, Germany\\
$^{8}$ School of Physics and Astronomy, University of Leicester, Leicester LE1 7RH, UK\\
$^{9}$ South African Astronomical Observatory, P.O Box 9, Observatory 7935, Cape Town, South Africa\\
$^{10}$ Department of Astronomy, University of Cape Town, Rondebosch 7700, Cape Town, South Africa\\
$^{11}$ Departement d'Astronomie, Universit\'e de Gen\`eve, 51 chemin Pegasi, 1290 Sauverny, Switzerland\\
$^{12}$Instituto de Astronom\'ia, Universidad Cat\'{o}lica del Norte, Casa Central, Angamos 0610, Antofagasta, Chile\\
$^{13}$Astronomy Unit, Queen Mary University of London, Mile End Road, London E1 4NS, UK\\
$^{14}$Astrophysics Group, Cavendish Laboratory, J.J. Thomson Avenue, Cambridge CB3 0HE, UK.\\
$^{15}$Astrophysics Research Centre, School of Mathematics and Physics, Queen's University Belfast, BT7 1NN Belfast, UK\\
$^{16}$Mullard Space Science Laboratory, University College London, Holmbury St Mary, Dorking, Surrey, RH5 6NT, UK\\
$^{17}$European Space Agency (ESA), European Space Research and Technology Centre (ESTEC), Keplerlaan 1, 2201 AZ Noordwijk, The Netherlands\\
$^{18}$Aix Marseille Univ, CNRS, CNES, LAM, Marseille, France\vspace*{-0.5cm}\\
}

\date{}

\pubyear{2022}

\begin{document}
\label{firstpage}
\pagerange{\pageref{firstpage}--\pageref{lastpage}}
\maketitle

\begin{abstract}
We report the discovery of \Nplanet\,, a massive hot Jupiter orbiting a low-mass star as part of the Next Generation Transit Survey (NGTS). The planet has a mass and radius of $2.36 \pm 0.21$\,\mjup\,and $1.33 \pm 0.03$\,\rjup, and an orbital period of 1.543 days. The host is a K3V ($T_{\rm eff}=4660 \pm 41$\, K) metal-poor (${\rm [Fe/H]}=-0.26 \pm 0.07$\, dex) dwarf star with a mass and radius of $0.72 \pm 0.04$\,\msun\,and $0.86 \pm 0.04$\,\rsun. Its age and rotation period of $10.02^{+3.29}_{-7.30}$\, Gyr and $17.88 \pm 0.08$\,d respectively, are in accordance with the observed moderately low stellar activity level.  When comparing \Nplanet\ with currently known transiting hot Jupiters with similar equilibrium temperatures, it is found to have one of the largest measured radii despite its large mass. Inflation-free planetary structure models suggest the planet's atmosphere is inflated by $\sim21\%$, while inflationary models predict a radius consistent with observations, thus pointing to stellar irradiation as the probable origin of \Nplanet's radius inflation. Additionally, \Nplanet's bulk density ($1.25 \pm 0.15$\,g/cm$^3$) is also amongst the largest within the population of metal-poor giant hosts ([Fe/H] < 0.0), helping to reveal a falling upper boundary in metallicity-planet density parameter space that is in concordance with core accretion formation models.  The discovery of rare planetary systems such as \Nstar\ greatly contributes towards better constraints being placed on the formation and evolution mechanisms of massive planets orbiting low-mass stars.
\end{abstract}
\clearpage
\section{Introduction}
\label{sec:intro}
The increasing number of planet discoveries has allowed us to classify exoplanets into distinct populations such as hot Jupiters (e.g, 51Peg b, \citeauthor{mayor1995jupiter}, \citeyear{mayor1995jupiter}; NGTS-2, \citeauthor{raynard2018ngts}, \citeyear{raynard2018ngts}), which are planets with orbital period P $<$ 10 days, and masses between 1$-$13 $\mjup$, ultra-short period (USP) planets, characterised by their P $<1$ day orbit \citep[e.g, Kepler-10b,][]{batalha2011kepler}, Neptune desert planets \citep[e.g, LTT9779b,][]{jenkins2019tess}, with masses and periods about 10$-$20\,M$_{\oplus}$ and P $<4$ days, respectively, and super-Earths \citep[e.g, Trappist-1,][]{gillon2017seven}, with M $<$10 M$_{\oplus}$. Amongst all planet populations, the hot and ultra-hot giant planets are the most likely to be detected given their relative proximity to the host star, which maximises the transit probability function and radial-velocity (RV) amplitudes. However, although giant planets are easily identified, observations show that their occurrence rates (f$_\text{p}$) around solar-type stars are about 10$\%$ \citep{cumming2008keck,hsu2019occurrence}, from which only $\sim 1 \%$ are hot Jupiters \citep{wright2012frequency}, whereas for low-mass stars, Jovian planets are even less common \citep{johnson2007new, bonfils2013harps}. 

Another key discovery that was made relatively early in the history of exoplanet studies, was the correlation between giant planets f$_\text{p}$ with stellar metallicities ([Fe/H]) \citep{gonzalez1997stellar,santos2001metal,fischer2005planet,2020MNRAS.491.4481O}, where not only such planets preferentially form around metal-rich stars but an increase in metallicity leads to a higher giant planet occurrence rates \citep{jenkins2017new,buchhave2018jupiter, barbato2019gaps}.  However, although the fraction of giant planets orbiting metal-poor stars ([Fe/H]$<$0.0~dex) is significantly lower than their more metal-rich counterparts, the fraction is far from zero.  In fact, a number of gas giants have been found orbiting stars with metallicities down towards an [Fe/H] of -0.5~dex.  \citet{mortier2012frequency} found that the fraction of gas giants orbiting stars in the metallicity range -0.7\--\,0.0 is actually $\sim$4\%, yet the hot Jupiters have a fraction below 1\%.  

The stellar mass also plays an important role in the types of planets that can be formed orbiting a specific type of star. \citet{johnson2010giant} found that higher mass stars tend to host more gas giant planets.  This result has been confirmed by other works \citep{reffert2015precise,jones2016four}, likely being explained by the relationship between host star mass and disc mass, whereby as the stellar mass decreases, and hence the disc mass decreases, there is less and less material with which to quickly form a giant planet before the disc disperses.  These results imply that metal-poor and low-mass stars should be relatively devoid of gas giant planets, particularly the short period hot Jupiter population.
\subsection{The Next Generation Transit Survey}
The Next Generation Transit Survey \citep*[NGTS;][]{Chazelas2012,McCormac2017,wheatley2018next} is a collection of 12 telescopes operating from the ESO Paranal Observatory in Chile, with the goal of detecting new transiting planetary systems. Each telescope has a diameter of 0.2 m, and with individual fields of view of 8 deg$^2$, a combined wide-field of 96 deg$^2$ can be obtained. Detectors are 2K $\times$ 2K pixels, with individual pixels measuring 13.5 $\mu$m, which corresponds to an on-sky size of 4.97 arcseconds, thus providing high sensitivity images over a wavelength domain between 520$-$890nm. This combination allows 150 ppm photometry to be obtained on bright stars ($V <$10~mags) for multi-camera observations, while for single telescope mode at 30min cadence, a precision of 400 ppm is achievable (Bayliss et al. 2022 in press). The project has been operational since February 2016, and over the past 6 years has so far acquired over 300 billion measurements of over 30 million stars. Within this treasure-trove of data, the NGTS has discovered 19 new planetary systems \citep*[e.g.][]{bayliss2018ngts, bryant2020ngts,tilbrook2021ngts}, with more yet to be confirmed. A few of the highlights include the discovery of the Neptune desert planet NGTS-4b \citep{west2019ngts}, an ultra short period Jupiter NGTS-6b \citep{vines2019ngts}, and the shortest period hot Jupiter NGTS-10b around a K5V star \citep{mccormac2020ngts}.  
Here we add to the success of this project by announcing the discovery of a new, massive hot Jupiter orbiting a low-mass star, \Nplanet.

The manuscript is organised as follows, in $\S$ \ref{sec:obs}, we present the photometry extraction from NGTS, TESS, and SAAO lightcurve and HARPS spectroscopic follow-up. $\S$ \ref{sec:analysis} describes the data analysis, where we extract stellar parameters ($\S$ \ref{sub:stellar}), assess TESS lightcurves dilution ($\S$ \ref{sub:TESSdilution}), and perform a global modelling to derive the planetary properties ($\S$ \ref{sub:globalmodeling}). Stellar rotation period as well as transit timing variation were probed in $\S$ \ref{sub:rotation} and \ref{sub:ttvmodeling}, respectively. Finally, we discuss our results in $\S$ \ref{sec:discussion} and set out the conclusions in $\S$ \ref{sec:concl}.
\section{Observations}
Here we describe the observation data reductions that led to the discovery of \Nplanet\,; Table \ref{tab:ngts} shows a portion of the photometry for guidance.
\label{sec:obs}
\subsection{\NGTS{} Photometry}
\label{sub:ngtsphot}
\begin{figure}
	\includegraphics[width=\columnwidth]{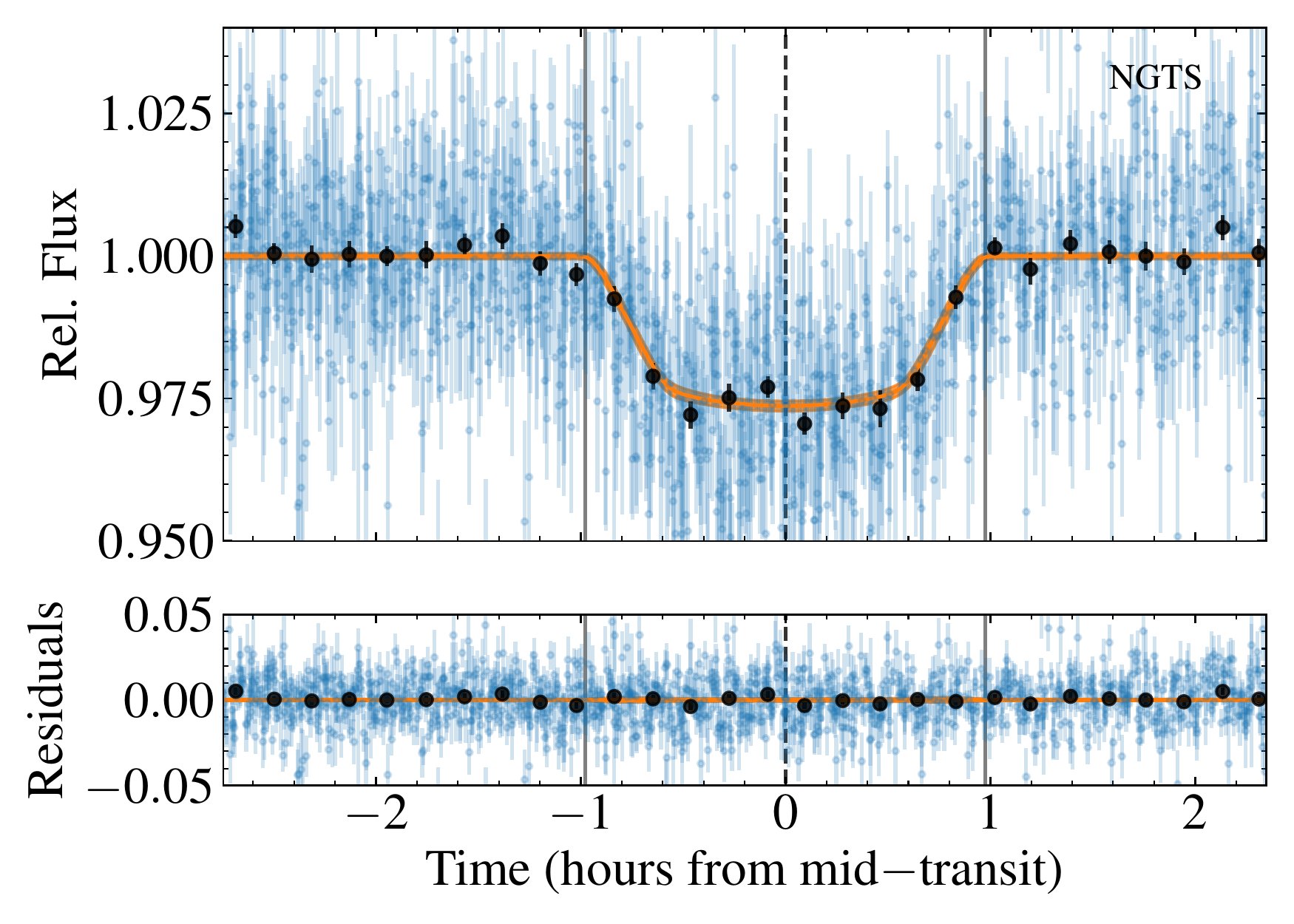}
    \caption{\textbf{Top}: NGTS detrended lightcurve phase-folded to the best-fitting period listed in Table \ref{tab:planet} and zoomed to show the transit event. Blue and black circles correspond to modelled photometric data and 11 min binned data with the associated photon noise error. The orange line and shaded region show the median transit model and its 1-$\sigma$ confidence interval. Solid gray vertical lines indicate the start of transit ingress (T$_{1}$) and end of egress (T$_{4}$). Dashed gray vertical line represents the transit centre (T$_{\rm c}$). \textbf{Bottom}: residuals to the best fit model.} 
    \label{fig:ngtsphot}
\end{figure}

\Nstar\ was observed during the 2018 campaign from March 24 to November 7, where 9157 images
were obtained during 150 nights, 
with 10s exposure time per frame. Prior to aperture photometry extraction with CASUTools\footnote{\url{http://casu.ast.cam.ac.uk/surveys-projects/software-release}} package, nightly trends such as atmospheric extinction were corrected for with an adapted version of the SysRem algorithm \citep{Tamuz2005}. Transit searches were carried out with our implementation of the box least-squares (BLS) fitting algorithm \citep[][]{kovacs2002box,collier2006fast} \texttt{ORION} code, where a total of 25 transits were detected, of which 13 were full transits. A strong signal was detected at 1.543 days, and a validation process began in order to either confirm the signal as a likely transiting hot Jupiter or reject it as a false positive detection. For example, background eclipsing binaries, where consecutive transits show odd-even and/or V shaped transits. \Nstar\ passed every validation step, and therefore further photometry and RV follow-up were obtained. Figure~\ref{fig:ngtsphot} shows the NGTS detection lightcurve wrapped around the best-fitting period $1.5433897 \pm 0.0000016$\,d computed from the global modelling ($\S$ \ref{sub:globalmodeling}). For a detailed description of the NGTS mission, data reduction, and acquisition, we refer the reader to \citet{wheatley2018next}.
\begin{table}
	\centering
	\caption{NGTS, TESS and SAAO photometry for \Nstar. The full Table is available in a machine-readable format from the online journal.  A portion is shown here for guidance.}
	\label{tab:ngts}
	\begin{tabular}{cccc}
	Time	&	Flux       &Flux & Instrument\\
    (BJD$_{\rm TDB}$-2457000)	&	(normalised)	&error & \\
	\hline
    ...  &   ...    &   ...  &  ...  \\
    1203.89922322& 1.0029& 0.0137& NGTS\\  
    1203.90264915& 1.0172& 0.0120& NGTS\\  
    1203.904506795& 1.0801& 0.0383& NGTS\\ 
    1204.872232445& 0.9932& 0.0083& NGTS\\ 
    1204.87575096& 0.9935& 0.0141& NGTS\\  
    1204.87920004& 0.9812& 0.0122& NGTS\\  
    1204.88266068& 1.0091& 0.0096& NGTS\\
    ...  &   ...    &   ...  &  ...  \\
    2051.58581& 0.9994& 0.0109& TESS\\ 
    2051.59276& 0.9968& 0.0109& TESS\\  
    2051.5997& 0.9674& 0.0110&  TESS\\  
    2051.60665& 0.9980& 0.0110&  TESS\\
    ...  &   ...    &   ...  &  ...  \\
    2051.43517437& 0.9811& 0.0083& SAAO\\ 
    2051.4358689& 0.9866& 0.0083& SAAO\\  
    2051.43656342& 0.9916& 0.0083& SAAO\\  
    2051.43725795& 0.9925& 0.0084& SAAO\\
        ...  &   ...    &   ... & ...  \\
	\hline
	\end{tabular}
\end{table}
%
\subsection{TESS Photometry}
\label{sub:tessphot}
\begin{figure}
	\includegraphics[width=\columnwidth]{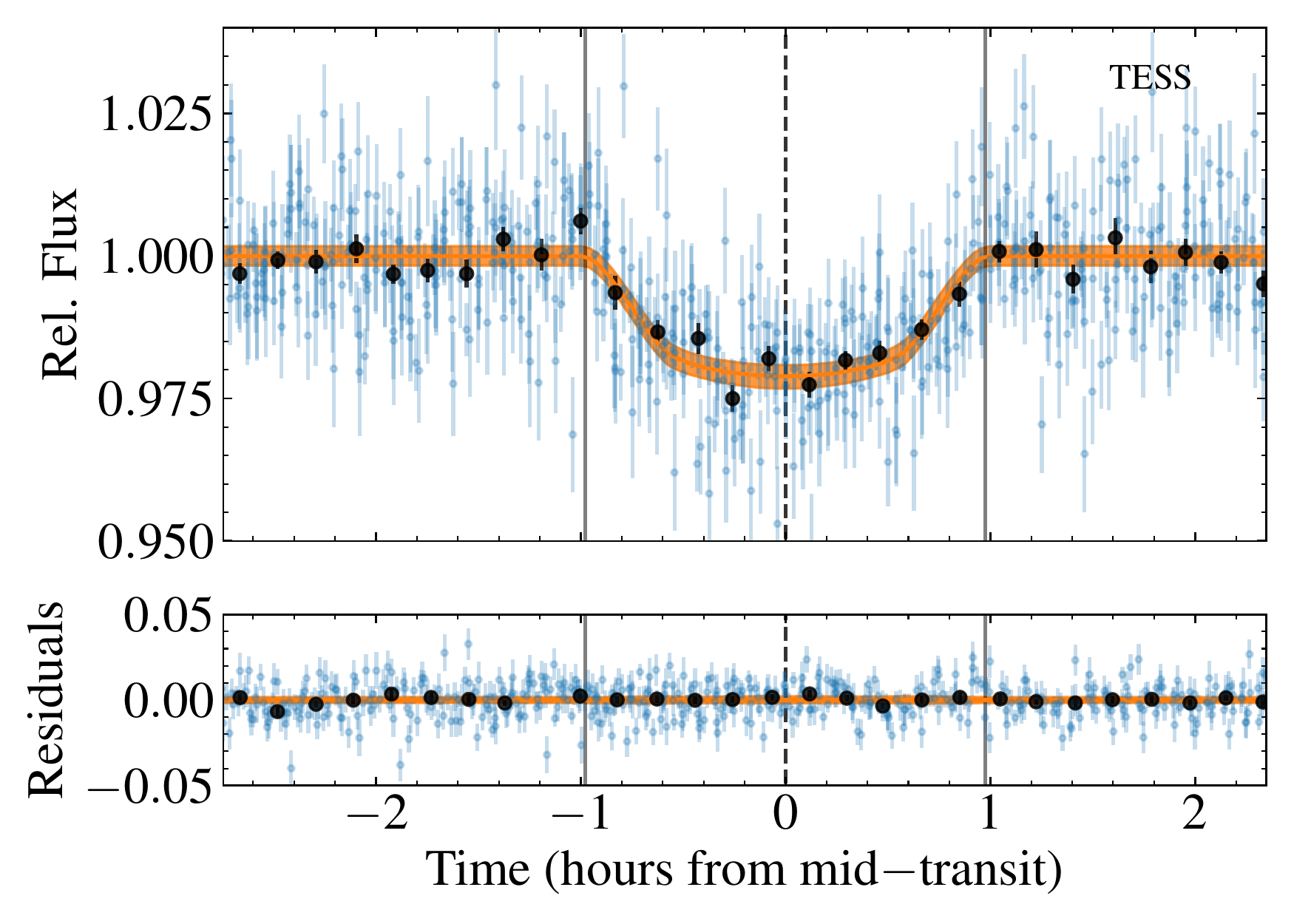}
    \caption{Phase-folded TESS detrended lightcurve. Coloured are labeled as in Fig. \ref{fig:ngtsphot}}
    \label{fig:tesslc}
\end{figure}

The Transiting Exoplanet Survey Satellite \citep[TESS;][]{TESS} observed \Nstar\ in sector 1 on universal time (UT) 2018-07-26 and sector 27 on UT 2020-7-5, with cadences of 30 and 10 minutes, respectively. Full Frame Images (FFIs) have been downloaded using the python astroquery module \citep{2019AJ....157...98G} to query the TESSCut service \citep{TESSCut}. For each sector a master image was calculated and used to both determine the star aperture and identify pixels for background correction, thus estimating \Nstar\ brightness for each image. Fig. \ref{fig:tesslc} shows the detrended phase-folded lightcurve, median, and 1-$\sigma$ best-fitting model from $\S$ \ref{sub:globalmodeling}. 
\subsection{SAAO Photometry}
\label{sub:saaophot}
\begin{figure}
	\includegraphics[width=\columnwidth]{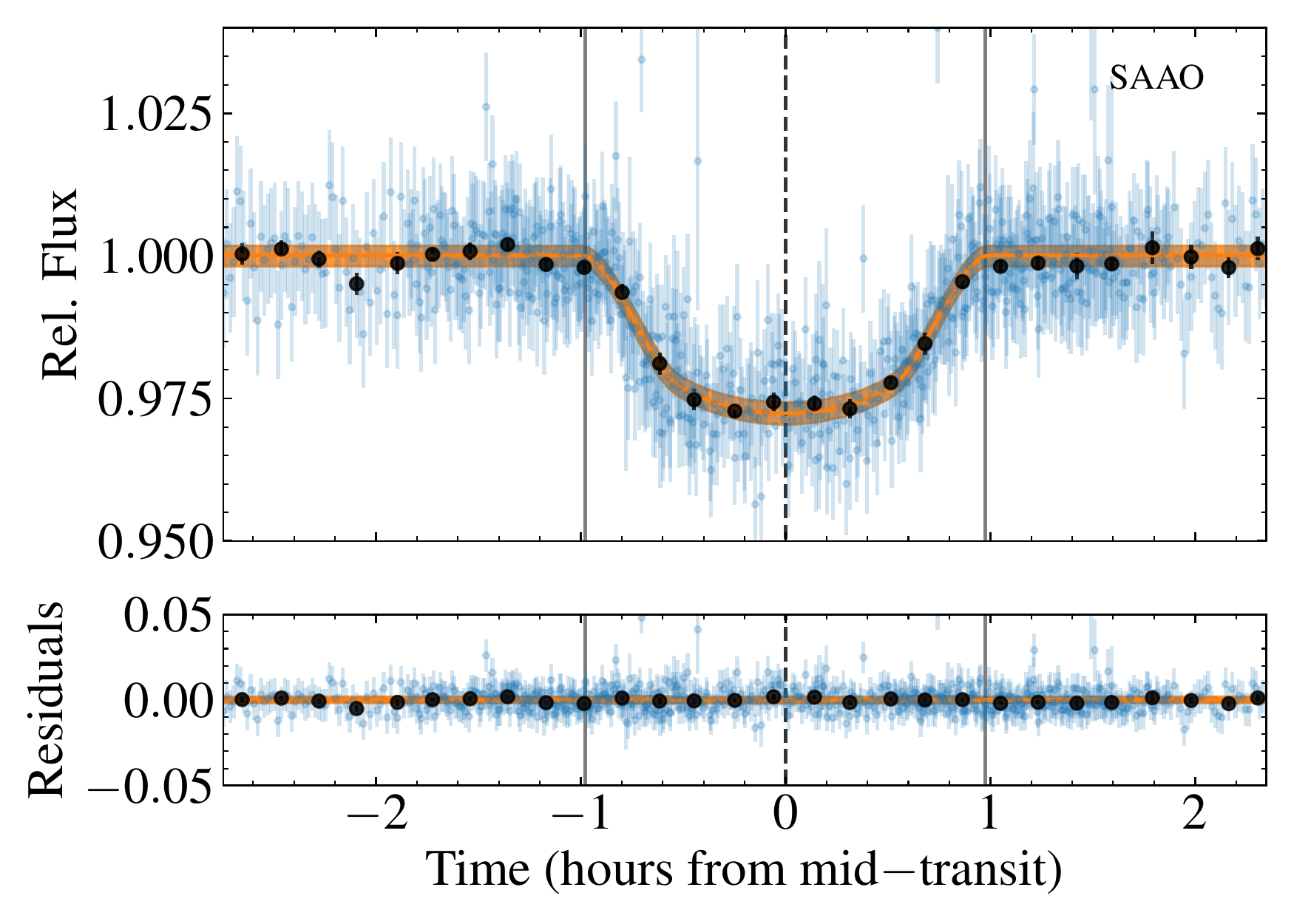}
    \caption{Phase-folded SAAO of the three detrended lightcurves. Coloured are labeled as in Fig. \ref{fig:ngtsphot}}
    \label{fig:saaolc}
\end{figure}
Follow up photometry of \Nstar\ was obtained with the South African Astronomical Observatory (SAAO) 1-m telescope equipped with the Sutherland High-Speed Optical Camera \citep[SHOC;][]{Coppejans2013}. The star was observed three times, on the nights of 2020 June 19\textsuperscript{th}, 2020 July 20\textsuperscript{th} and 2020 July 23\textsuperscript{rd}. All observations were taken in $V$ band with 60 seconds of exposure times. 

The data were reduced using the \texttt{safphot}\footnote{https://github.com/apchsh/SAFPhot}, a custom python package for the reduction of SAAO photometric data. Standard flat field and bias corrections were applied by \texttt{safphot}, which then utilises the \texttt{sep} package \citep{Barbary2016} to extract aperture photometry for both the \Nstar\ and nearby comparison stars with which to perform differential photometry. \texttt{sep} also measured and subtracted the sky background, while masking the stars in the image and adopting box sizes and filter widths that minimised residuals across the frame. Two nearby, bright comparison stars were used to perform differential photometry, with aperture sizes ranging between 4.4 and 5.9 pixels for the target dependent on the seeing level.

The July 2020 observations both captured complete transits of \Nplanet, whereas the 2020 June 19\textsuperscript{th} was affected by clouds during mid-transit, thus leading to gaps at that portion. However, since both ingress and egress were observed, the data was used in the global modelling in $\S$ \ref{sub:globalmodeling}, where Fig. \ref{fig:saaolc} shows the phase-folded SAAO follow-up lightcurve.
\subsection{Spectroscopic Follow Up}
\label{sub:spect}
\begin{figure}
	\includegraphics[width=\columnwidth]{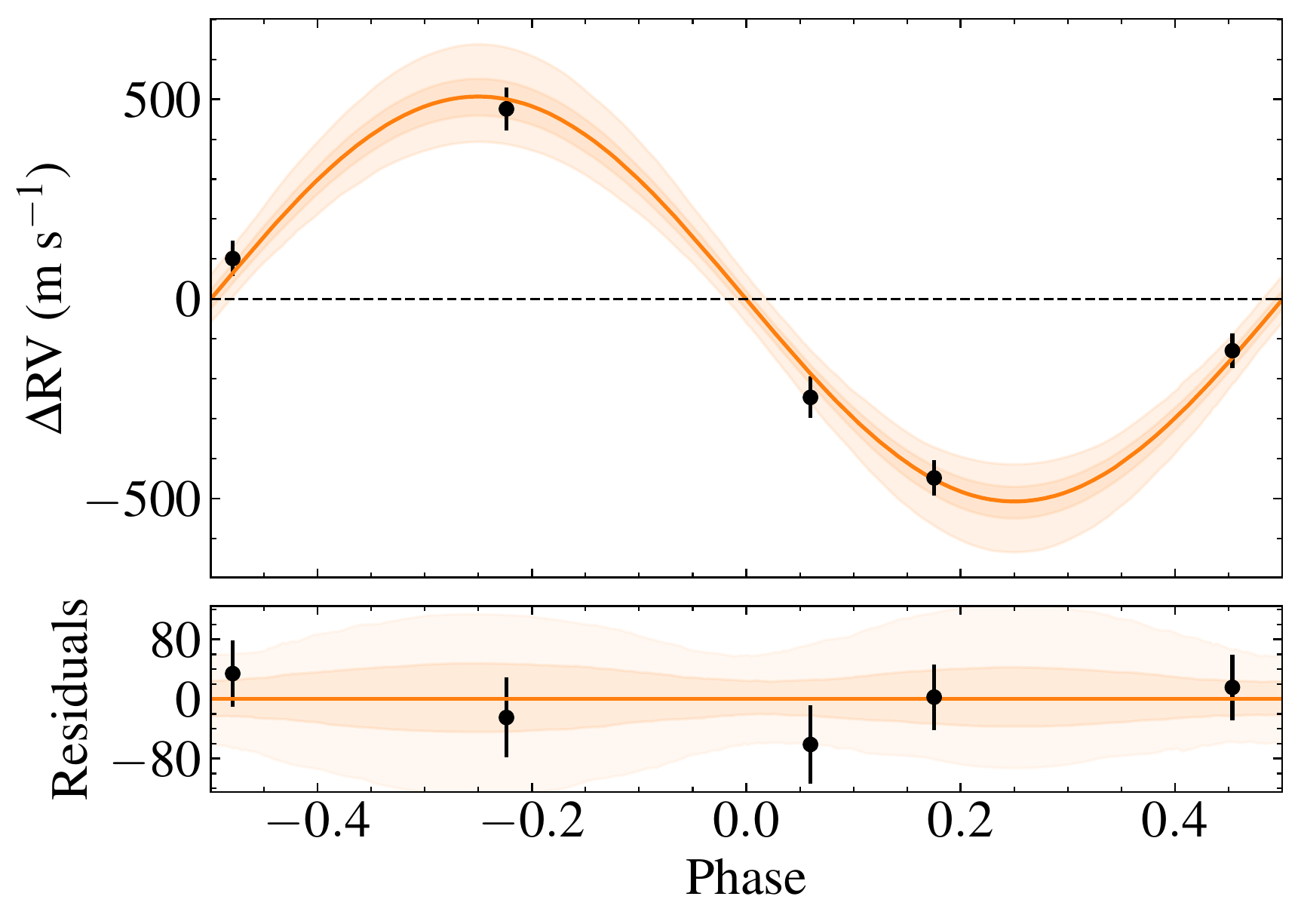}
    \caption{\textbf{Top}: Phase-folded HARPS radial-velocity shown in black, with its 1- and 2-$\sigma$ confidence intervals in shades of orange. \textbf{Bottom}: residuals to the best fit with RMS$_\text{RV} = 33.75~ \text{m s}^{-1}$}
    \label{fig:harps}
\end{figure}

Five high-resolution spectra for \Nstar\ were obtained during UT 16-07-2021 and 07-09-2021 under the HARPS prog ID (Wheatley  0105.C-0773) on the ESO 3.6\,m \citep{2003Msngr.114...20M} telescope at the la silla observatory in Chile. Due to the apparent faintness of the star and large expected RV amplitude, we used HARPS in the high efficiency mode (EGGS), which trades resolution for high throughput. The EGGS science fibre is 1.4 arcseconds when projected on-sky, which with exposure times of 2400$-$2700 seconds, we achieved a signal-to-noise (SNR) of 4$-$5 per pixel at 5500\,\AA. The RV measurements were computed with the standard HARPS pipeline using the following binary masks for the cross-correlation: G2, K5, K0, and M4, where agreement was found amongst the RVs estimated with these binary masks. Therefore, given \Nstar\ spectral type, we adopted the RV data estimated with the binary K5 mask, which is shown, accompanied by the best-fitting Keplerian model, in Fig. \ref{fig:harps} as well as in Table \ref{tab:rvs} with additional diagnostics data.

\begin{figure}
	\includegraphics[width=\columnwidth]{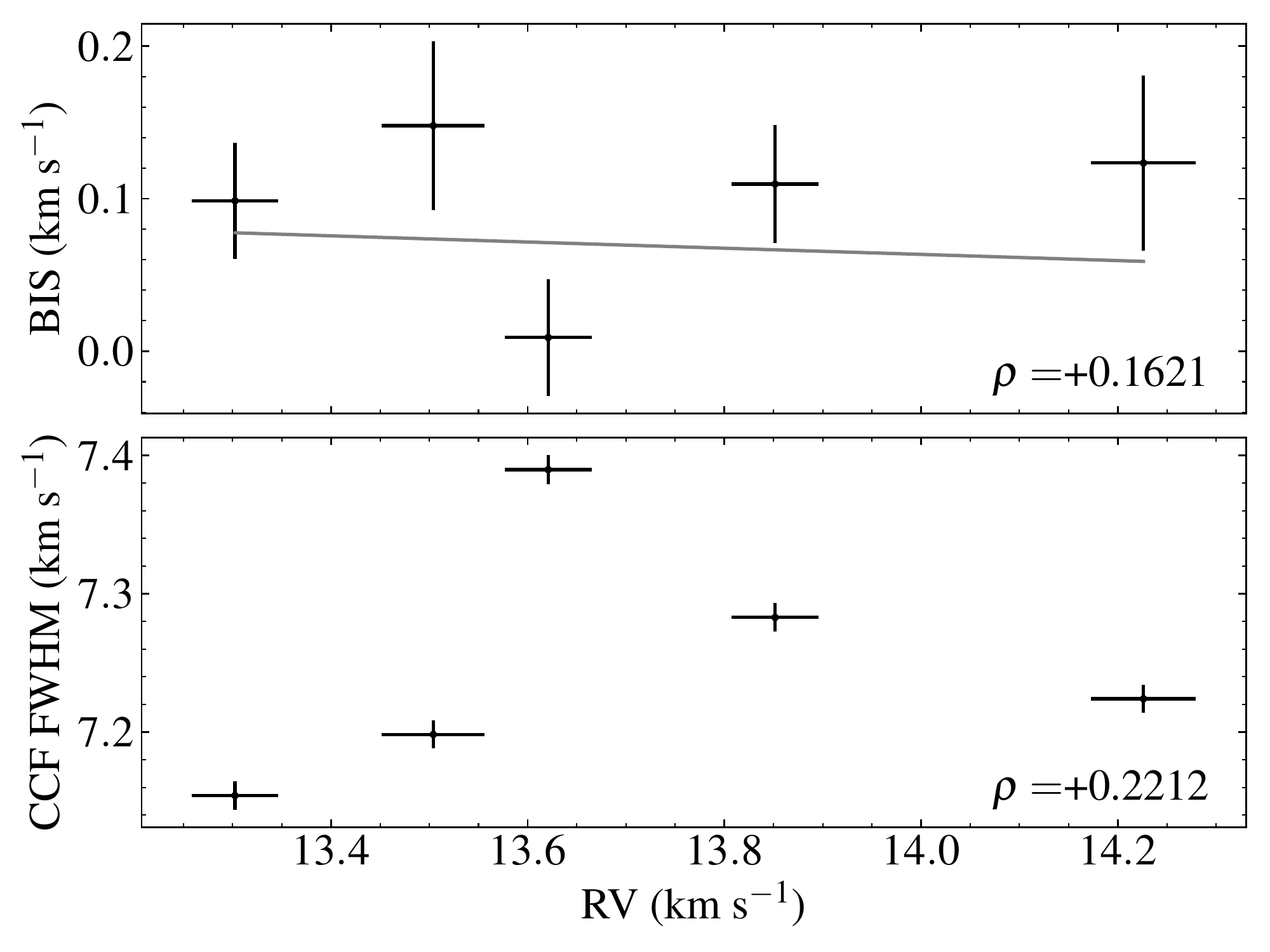}
    \caption{\textbf{Top}: radial-velocity vs bisector span measurement with a best-fitting linear model shown as gray solid line. The Pearson r coefficient supports no significant correlation between datasets, with $\rho$ shown at the lower right. \textbf{Bottom}: RV against CCF full width at half maxima.}
    \label{fig:bisccf}
\end{figure}

Since stellar activity has long been recognised to mimic planetary signals, we investigated whether correlations are present between activity diagnosis parameters to rule out the possibility of a false positive signal. Figure \ref{fig:bisccf} upper panel shows the RV measurements vs bisector velocity span (BIS) of the cross-correlation function (CCF) with a best-fitting linear model. A Pearson r coefficient, which measures the correlation between datasets, approaches zero ($\rho=$+0.1621), thus pointing to negligible correlation between the RVs and BIS. The full width at half maxima (FWHM) is also shown in the bottom panel indicating no trend between RV and CCF$-$FWHM. Finally, the lightcurves low rotational modulation and lack of observable flares in the lightcurves are in accordance with a moderately quiet star, thus supporting the RV and transit detected signal as coming from \Nplanet.

Due to variations of up to 236 ${\rm ms^{-1}}$ in the CCF$-$FWHM (Fig. \ref{fig:bisccf}), likely caused by the very low SNR, we performed modelling tests based on $\S$ \ref{sub:globalmodeling} to investigate whether the removal of the second HARPS spectrum listed in Table \ref{tab:rvs} would impact the posterior distributions derived when the entire RV data is included in the model. Since the tests yielded posterior distributions that are in strong statistical agreement, we included every RV measurement while building our global model in $\S$ \ref{sub:globalmodeling}.
\begin{table}
	\centering
	\caption{HARPS follow-up Radial Velocities for \Nstar}
	\label{tab:rvs}
	\tabcolsep=0.11cm
	\begin{tabular}{cccccc} 
BJD$_{\rm TDB}$		&	RV		&RV err &	FWHM& 	BIS\\
-2457000	& (\ms)& (\ms)&(\ms) &(\ms) \\
		\hline
2411.818920&	13851&	27&	7283&		110\\
2428.692090&	13620&	27&	7390&		009\\
2460.673898&	13302&	27&	7154&		099\\
2463.582604&	13503&	39&	7198&		148\\
2464.688277&	14226&	41&	7224&		123\\
		\hline
	\end{tabular}
\end{table}
\section{Data Analysis}
\label{sec:analysis}
\subsection{Stellar Properties}
\label{sub:stellar}
\Nstar\ properties were independently derived using the packages, spectroscopic parameters and atmospheric chemistries of stars \citep[\texttt{SPECIES}\footnote{github.com/msotov/SPECIES};][]{soto2018spectroscopic} and the spectral energy distribution Bayesian model averaging fitter \citep[\texttt{ARIADNE} \footnote{https://github.com/jvines/astroARIADNE};][]{vines2022ariadne}.

\texttt{SPECIES} estimates atmospheric parameters such as effective temperature (T$_{\rm eff}$), [Fe/H], surface gravity ($\log g$), and microturbulence velocity ($\xi_t$) from high resolution spectra. First, \texttt{SPECIES} computes the equivalent widths ($W$) of Fe I and Fe II lines with the \texttt{ARES} code \citep{sousa2007new}. An appropriate atmospheric grid of models computed from interpolating ATLAS9 \citep{castelli2004new} atmosphere model as well as $W$ are handed to MOOG \citep{sneden1973nitrogen}, which solves the radiative transfer equation (RTE) while measuring the correlation between Fe line abundances as a function of excitation potential and $W$, assuming Local thermodynamic equilibrium (LTE). While solving the RTE, the correct atmospheric parameters are determined through an iterative process carried out until no correlation is found between the iron abundance with the excitation potential, and with the reduced equivalent width ($W/\lambda$). The Mass, radius and age are obtained from the \texttt{isochrone} package \citep{morton2015isochrones} by interpolating through a grid of MIST \citep{dotter2016mesa} evolutionary tracks with T$_{\rm eff}$, [Fe/H],  $\log g$ priors previously derived as well as parallax, photometry in several bands, and proper motions. Nested sampling \citep[][]{feroz2009multinest} is used to estimate posterior distributions for \mstar,\,\rstar, and age. Rotation and macro turbulent velocities are calculated from temperature calibrators and fitting the absorption lines of observed spectra with synthetic line profiles. From our analysis using \texttt{SPECIES} we derived the following stellar properties with their 1-$\sigma$ confidence interval, T$_{\rm eff} = 4746 \pm 300$\,K, $\text{[Fe/H]} = -0.26 \pm 0.09$\,dex, $\log \text{g} = 4.57 \pm 0.45$, $\text{Age} = 11.72^{+1.28}_{-2.36}$\,Gyr, $\mstar = 0.75 \pm 0.01~ \msun$, and $\rstar = 0.74 \pm 0.01~ \rsun$. Chemical abundances and $vsini$ were not extracted due to the low SNR achieved at this faint regime (V=15.6).

\begin{figure}
	\includegraphics[width=\columnwidth]{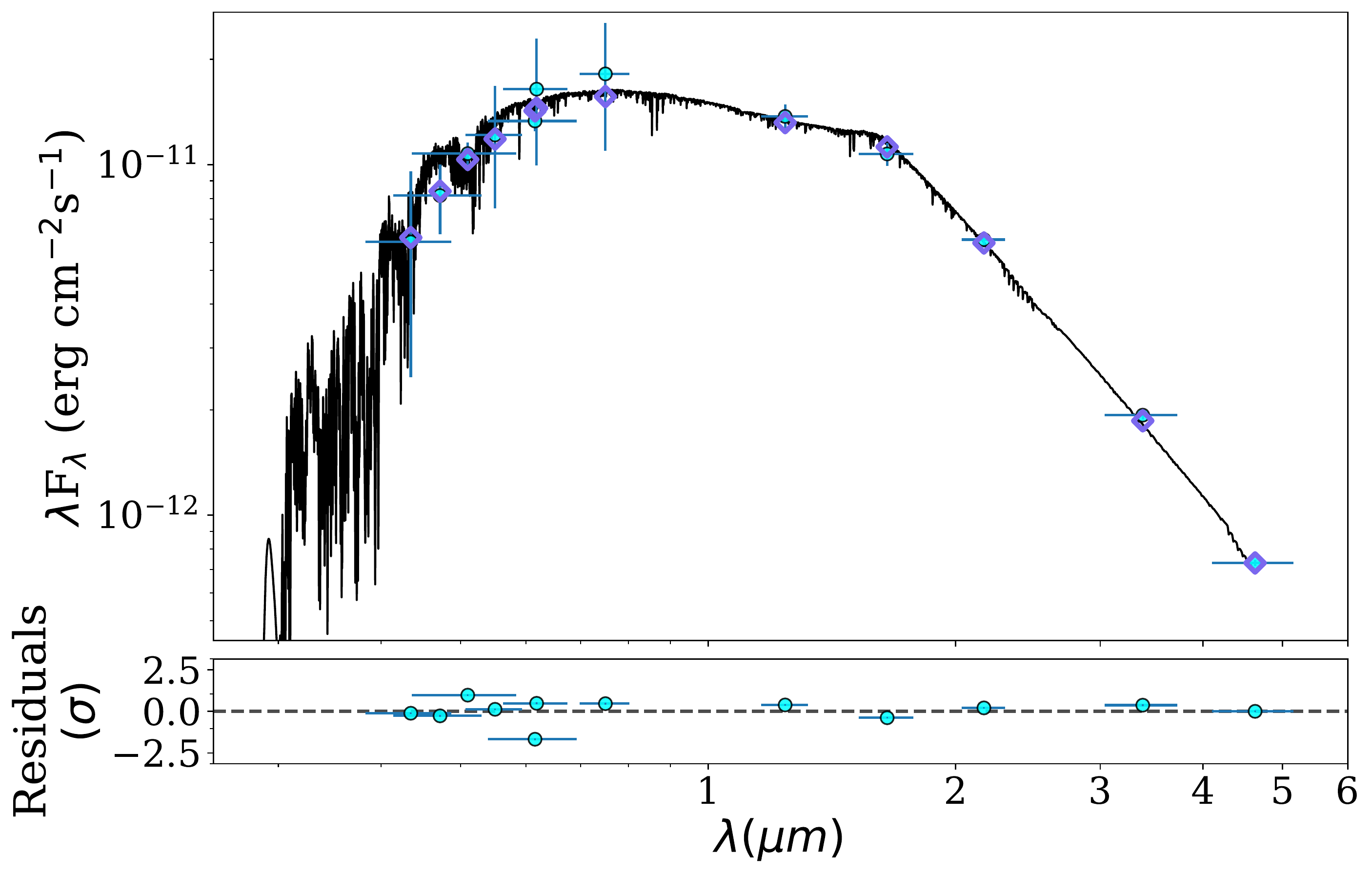}
    \caption{\textbf{Top}: The best-fitting spectral energy distribution (black line) based on \citet{castelli2004new} given the \Nstar\ photometric data (cyan points) and their respective bandwidths shown as horizontal errorbars. Purple diamonds represent the synthetic magnitudes centred at the wavelengths of the photometric data from Table~\ref{tab:stellar}. \textbf{Bottom}: residuals to the best fit in $\sigma$ units.}
    \label{fig:sed}
\end{figure}

We have also estimated \Nstar\ parameters using the \texttt{ARIADNE} python package \citep{vines2022ariadne}, which is an automated code that extract stellar parameters by fitting archival photometry to different stellar atmosphere models using Nested Sampling through \texttt{DYNESTY} \citep[][]{speagle2020dynesty}. The \texttt{SPECIES} derived stellar properties T$_{\text{eff}}$, $\log g$, and [Fe/H] as well as archival photometric data were used as \texttt{ARIADNE} input to fit the spectral energy distribution (SED) using different models (Fig. \ref{fig:sed}). These models were convolved with several filter response functions \citep[see available SED models in][]{vines2022ariadne}, where synthetic fluxes scaled by $(R/D)^{2}$ were estimated from interpolating through the model grids, which are functions of T$_{\text{eff}}$, $\log g$, [Fe/H], and $V$ band extinction (A$_{\text{V}}$).
An excess noise parameter is modelled for each photometric measurement to account for underestimated uncertainties. The final stellar parameters are derived from the averaged posterior distributions from the Phoenix V2 \citep{husser2013new}, BT-Settl \citep{hauschildt1999nextgen,allard2012models}, \citet{castelli2004new}, and \citet{kurucz1993atlas9} SED models, weighted by their respective Bayesian evidence estimates. \texttt{ARIADNE} parameters T$_{\text{eff}}$, $\log g$, [Fe/H] as well as additional quantities such as distance, stellar radius, and A$_{\text{V}}$ were used to derive stellar age, mass and the equal evolutionary points from the \texttt{isochrone} package.
Table \ref{tab:stellar} shows the adopted stellar properties from \texttt{ARIADNE}, which due to its Bayesian averaging method computed precise stellar parameters, particularly the \rstar and T$_{\rm eff}$, which were key to inform the global modelling of \Nplanet\ (see $\S$ \ref{sub:globalmodeling}). 

For consistency, we compared the GAIA DR3 stellar parameters T$_{\rm eff}$ = 4665$^{+15}_{-24}$\,K, log g = 4.53$^{+0.03}_{-0.07}$\,dex, \rstar\ = 0.83$^{+0.12}_{-0.04}$\,\rsun\,, and distance of 612$^{+85}_{-27}$\,pc, with both \texttt{SPECIES} and \texttt{ARIADNE}, and found the measurements to be in statistical agreement. The GAIA astrometric excess noise as well as the renormalised unit weight error are 0 and 1.005, respectively, which are consistent with \Nstar\ being a single star system.
\subsubsection{Age Estimation}
\label{subsub:Agestimation}
The age of stars are commonly estimated using grids of pre-computed stellar evolutionary models described by stellar physical properties (e.g, temperature, luminosity, metallicity, etc.) that are interpolated to fit a set of observed stellar parameters. Such evolutionary models could be rearranged to tracks of fixed ages, i.e, isochrones, from which stellar ages are estimated. However, the complexity and strong non-linearity of isochrones along with observational uncertainties make it difficult to precisely estimate stellar ages. 

Although \texttt{ARIADNE} and \texttt{SPECIES} show consistent ages posteriors, the former gives a broader distribution than the latter. Therefore, we assessed \Nstar\ age based on gyrochronology models, which assume that stellar ages are a first order function of the rotation period, thus relying on less assumptions compared to other age estimation methods. The gyrochronology models we used were based on \citet{barnes2007ages, mamajek2008improved}, and \citet{meibom2009stellar}, which point to an age between 1 and 4.5 Gyr for a rotation period (P$_{\rm rot}$) of about 18 days (see $\S$ \ref{sub:rotation} for the P$_{\rm rot}$ calculation). Additionally, we used the \texttt{stardate} \citep{angus2019toward} code, which combines the \texttt{isochrone} package with gyrochronology models, thus computing an age of $4.94^{+3.56}_{-2.59}$\,Gyr.
Since pure gyrochronology models as well as the joint analysis with isochrone fitting yield ages in statistical agreement with \texttt{ARIADNE}, we adopted the \texttt{ARIADNE} age of $10.0^{+3.29}_{-7.30}$\, Gyr. Yet, \Nstar\ age lower end may be more likely given its moderately low activity supported by its lack of flares as well as measured rotational period and amplitude ($\S \ref{sub:rotation}$).   

Finally, we checked \Nstar\ spectrum for lithium lines, which due to its volatility with temperature, its abundance are depleted quickly in stellar atmospheres already in the first hundred million years of the star lifetime, hence the existence of photospheric Li is frequently associated to young stars \citep[e.g, see][]{christensen2018ages}. Therefore, we searched for Li lines in the averaged spectra, particularly around the strong Li resonant doublet at 6708 \AA, and found no evidence for Li lines, thus giving further constraints in \Nstar\ lower age limit (> 50-100 Myr).
\begin{table}
	\centering
	\caption{Stellar Properties for \Nstar}
	\begin{tabular}{lcc} 
	Property	&	Value		&Source\\
	\hline
    \multicolumn{3}{l}{Astrometric Properties}\\
    R.A.		&	\mbox{$20^{\rmn{h}} 45^{\rmn{m}} 01\fs9941$}			&GAIA\\
	Dec			&	\mbox{$-35\degr 25\arcmin 40\farcs 2322$}			& GAIA	\\
	2MASS I.D.	& J20450201-3525401	&2MASS	\\
	TIC I.D.	& 441422655	&TIC	\\
	GAIA DR3 I.D. & 6779308394419726848	&GAIA	\\
	Parallax (mas) & 1.71 $\pm$ 0.03 &GAIA \\
    $\mu_{{\rm R.A.}}$ (\masy) & -13.443 $\pm$ 0.031 & GAIA \\
	$\mu_{{\rm Dec.}}$ (\masy) & -7.834 $\pm$ 0.028 & GAIA \\
    \\
    \multicolumn{3}{l}{Photometric Properties}\\
	V (mag)		&15.621 $\pm$ 0.096 &APASS\\
	B (mag)		&16.648 $\pm$ 0.107	&APASS\\
	g (mag)		&16.108	$\pm$ 0.048	&APASS\\
	r (mag)		&15.241 $\pm$ 0.076	&APASS\\
	i (mag)		&14.856 $\pm$ 0.203	&APASS\\
    G (mag)		&15.22400 $\pm$ 0.00041		& GAIA\\
    NGTS (mag)	&14.82		&This work\\
    TESS (mag)	&14.5499 $\pm$ 0.006	&TIC\\
    J (mag)		&13.622 $\pm$ 0.027		&2MASS	\\
   	H (mag)		&13.105 $\pm$ 0.028	&2MASS	\\
	K (mag)		&12.951 $\pm$ 0.029	&2MASS	\\
    W1 (mag)	&12.898 $\pm$ 0.024	&WISE	\\
    W2 (mag)	&12.969 $\pm$ 0.027	&WISE	\\
    W3 (mag)	&12.485 $\pm$ 0.515	&WISE	\\
    
    \\
    \multicolumn{3}{l}{Derived Properties}\\
    $\rho_{*}$ (\gccc)   & 1.62 $\pm$ 0.10          &\texttt{Juliet}\\
    $\gamma_{RV}$ (\kms) & 13.75 $\pm$ 0.02     &\texttt{Juliet}\\
    P$_{\rm rot}$ (days) &  17.89 $\pm$ 0.08 & This work\\

    T$_{\rm eff}$ (K)    & 4660	 $\pm$ 41       &\texttt{ARIADNE}\\
    $[$Fe/H$]$	 & -0.26 $\pm$ 0.07       &\texttt{ARIADNE}\\
    log g                &	4.63 $\pm$ 0.34  &\texttt{ARIADNE}\\
    Age	(Gyr)		     & 10.02 $^{+3.29}_{-7.30}$  &\texttt{ARIADNE}\\
    \mstar (\msun)       & 0.72 $\pm$ 0.04          &\texttt{ARIADNE}\\
    \rstar (\rsun)       & 0.86 $\pm$ 0.04 	      &\texttt{ARIADNE}\\
    Distance (pc)	     & 640.98$^{+26.96}_{-23.59}$   &\texttt{ARIADNE} \\
	\hline
    \multicolumn{3}{l}{2MASS \citep{2MASS}; TIC v8 \citep{stassun2018tess};}\\
    \multicolumn{3}{l}{APASS \citep{APASS}; WISE \citep{WISE};}\\
    \multicolumn{3}{l}{{\em Gaia} \citep{brown2021gaia}}\\
	\end{tabular}
    \label{tab:stellar}
\end{table}
\subsection{Assessment of TESS lightcurve dilution}
\label{sub:TESSdilution}
TESS lightcurves are very susceptible to dilution, particularly in crowded fields where several contaminants may be within a few arcseconds from the target star. Due to its large plate scale of 21"/pixel, nearby stars may fall inside the photometric aperture, thus causing blends that affect transit depth, which in turn underestimate fundamental planetary properties such as planet radius and bulk density. To account for this, we assess the level of contamination in TESS lightcurves by estimating the dilution factor (D) to be used as a prior in the global modelling.

\begin{figure}
	\includegraphics[width=\columnwidth]{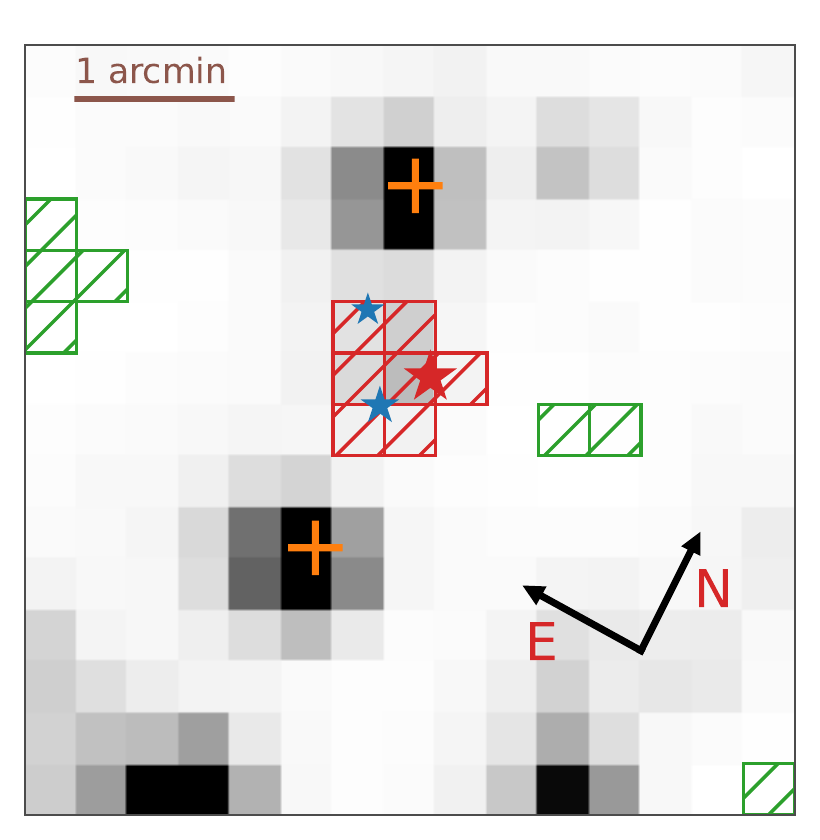}
    \caption{TESS Sector 1 Full-Frame Image cutout (15 x 15 pixels) centered at \Nplanet\ (red star). Red squares represent the mask used for the photometry extraction, green squares are the pixels utilised to estimate the sky background brightness. Stars marked in blue represent TIC-441422661 with a V=18.2 (upper left) and TIC-441422652 with V=17.3 (lower left). Orange pluses show the brightest stars in the field, TIC-441422643 (V=13.4) and TIC-441422667 (V=14) at the bottom left and upper right, respectively}.
    \label{fig:tesscutfile}
\end{figure}

A comparison between transit depths from the three missions shows that the planet-to-star radius ratio estimated from NGTS ($0.161 \pm 0.003$) and SAAO ($0.157 \pm 0.003$) are consistent, whilst TESS smaller ratio ($0.138 \pm 0.005$) indicates a shallower transit depth likely caused by two stars in the photometric aperture (see Fig. \ref{fig:tesscutfile}). To estimate the dilution level, we used the \texttt{ARIADNE} code to compute the SED for \Nstar\ and the two contaminants, TIC-441422661 and TIC-441422652, which are 37.86 and 22.61 arcseconds from \Nstar,respectively. Synthetic fluxes were computed by \texttt{ARIADNE}, where a theoretical D $\sim 20.5 \%$ in the TESS band, from Eq. \ref{eq:dilutioneq}, where $F_{cont}$ and $F_{target}$ represent the contaminant and target fluxes, respectively.
\begin{equation}
    D = \sum_{cont}{F_{cont}/F_{target}}
    \label{eq:dilutioneq}
\end{equation}
For consistency, we have also estimated the dilution directly from the phase-folded lightcurves transit depths offset between TESS and NGTS. We assumed no dilution for the later since the contaminants light contribution inside the NGTS apertures are, if any, negligible due to their relative distances ($>22.6"$) to \Nstar\ as well as their faint magnitudes (V$>17$ mag). We found a dilution of $\sim 26.5\, \%$, which is 6$\%$ larger than the predicted dilution from the SED fitting. This may be due to some fractional flux entering the aperture coming from the two brighter stars in Fig. \ref{fig:tesscutfile}, which are flagged with oranges crosses. Upon running several tests by varying the dilution prior distribution, we chose a Gaussian prior centred at $\mu = 28\ \%$ and $\sigma = 10\%$ for the global modelling (see $\S$ \ref{sub:globalmodeling}), thus resulting in a posterior dilution D=$27.5^{+6.1}_{-5.9}\, \%$.
\subsection{Global modelling}
\label{sub:globalmodeling}
We performed a joint radial-velocity and photometric analysis with the \texttt{Juliet} \citep{espinoza2019juliet} python package, which is a versatile code wrapped around the \texttt{Batman} \citep{kreidberg2015batman} for lightcurve modeling and \texttt{radvel} \citep{fulton2018radvel} for RV analysis. Our dataset consists of five HARPS RVs and a total of 13616 photometric data points from NGTS, TESS and SAAO.

\begin{figure*}
	\includegraphics[width=2\columnwidth,angle=0]{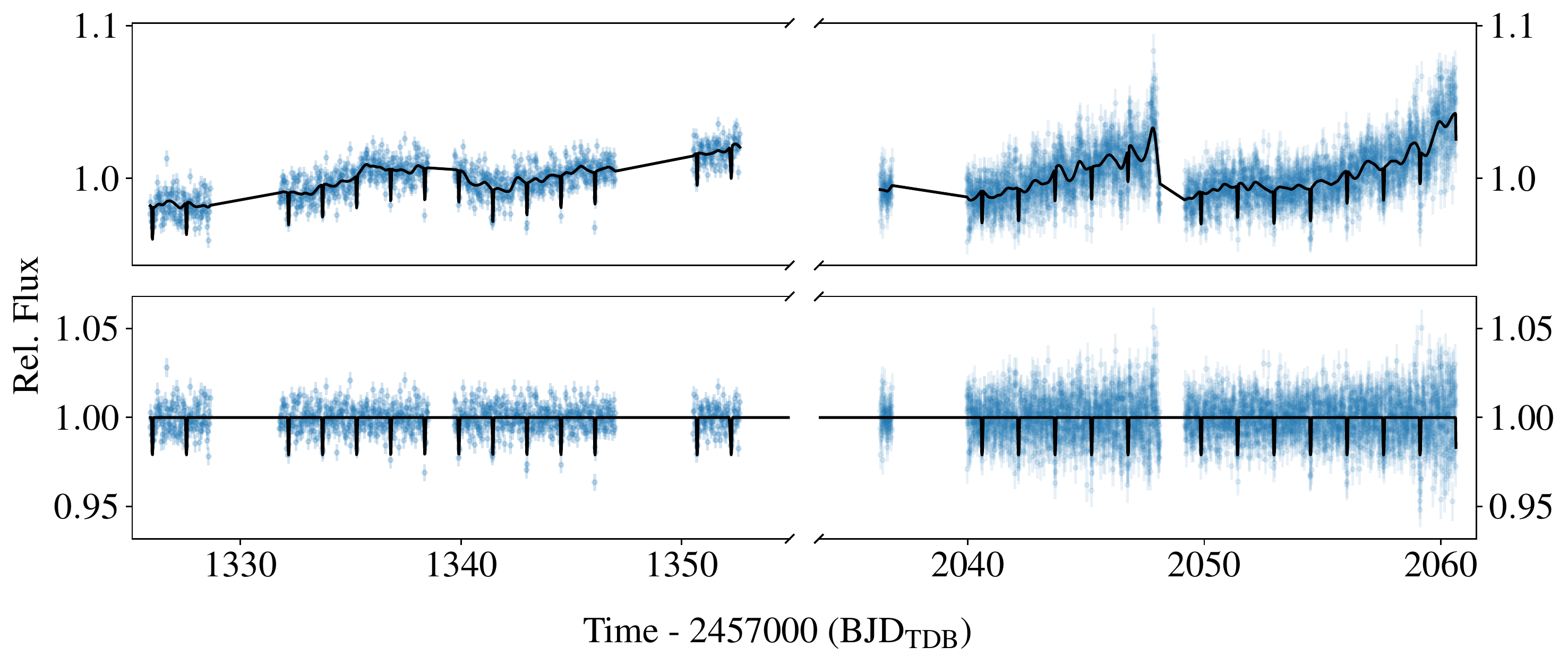}
    \caption{\textbf{Top}: TESS lightcurve extracted from FFIs showing moderate systematic (in blue), and its joint transit and Gaussian Process best-fitting median model in black. Left and right y-axis correspond to 30 and 10 min cadence lightcurves from Sector 1 and 27, respectively. \textbf{Bottom}: Detrended lightcurve in blue with its transit-only best-fitting model in black.}
    \label{fig:tesslcDetrendedVsnot}
\end{figure*}

Since each instrument has its own precision, and work under distinct environmental conditions, each dataset encapsulates noise differently, thus requiring a proper modelling so that planetary properties are optimally derived. For that reason, we included Gaussian processes (GP) in the noise model to account for correlated noise in the lightcurves, where each instrument was modelled by an approximate Matern Kernel. No GP was added to the Keplerian part of the global model due to the risk of overfitting caused by the low number of RV points. Although a global GP kernel is preferred for a proper modelling of stellar activity, we use a multi-instrument GP approach because \Nstar\ presents moderately low activity compared to instrumental systematic, particularly the TESS lightcurve, which shows the largest correlated noise (see Fig. \ref{fig:tesslcDetrendedVsnot} a) amongst the dataset. 

As discussed in $\S$ \ref{sub:TESSdilution}, TESS photometry is diluted by at least two contaminants, therefore a dilution normal prior ($\mathcal{N}(0.78, 0.1^2)$) was added specifically for this instrument\footnote{\texttt{Juliet} dilution definition is given by $\frac{1}{1+D}$, with $D$ from Eq. \ref{eq:dilutioneq}}, whereas NGTS and SAAO lightcurves had dilution factors set to undiluted.

For the limb darkening, we used the approach described in \citet{kipping2013efficient}, where a quadratic parametrisation with $q_1$ and $q_2$ using
uniform priors $\mathcal{U}(0, 1)$ were introduced for each instrument. The eccentricity $e$ was fixed to zero due to the small number of RV points. Yet, this assumption is supported by (1) observations of short period HJs (P < 4d), which are frequently found in circular orbits, and (2) \Nplanet\ tidal circularisation timescale $\tau$ of $\sim$ 1-11 Myrs, which was computed with Eq. 3 from \citet{adams2006long}, assuming a tidal quality factor Q$_p$ of $10^5-10^6$. Such a short $\tau$ compared to the planetary system lifetime quickly circularised the planet's orbit through planet-star dynamical interactions. Although a circular orbit was adopted, we ran tests to investigate whether a model with free $e$ would be preferred based on the Bayesian information criterion (BIC). The BIC is a model selection tool useful to test whether an increase in likelihood justifies the addition of new parameters in the tested model, which in turn, could lead to overfitting. The runs with non-circular orbits provided an upper limit of $e<0.12$ at 1$-\sigma$ and a BIC of 27.1, while the run with circular orbit yielded a BIC of 25.5. Therefore, given that models with lower BIC values are  favoured, the \Nstar\ global modelling with a circular orbit was preferred.

The radial-velocity part of the global model includes a Keplerian, a systemic RV term ($\gamma_{RV}$) and a white noise term to account for stellar jitter. Finally, due to the high dimension of the parameter space, we used the dynamic nested sampling algorithm \citep{higson2019dynamic} through \texttt{DYNESTY} with 1,000 live points.
\begin{table}
	\centering
	\caption{Planetary Properties for \Nplanet}
	\begin{tabular}{lc} 
	Property	&	Value \\
	\hline
    P (days)		        & 1.5433897 $\pm$ 0.0000016	\\
	T$_C$ (BJD$_{\rm TDB}$)		& 2458228.77853 $\pm$ 0.00067	 \\
    T$_{14}$ (hours) & 1.95 $\pm$ 0.03 \\
    $a/\rstar$		        & 5.89 $\pm$ 0.12            \\
    \rpl/\rstar & 0.159 $\pm$ 0.003 \\
    $b$ & 0.63 $\pm$ 0.03                       \\
    $i(deg)$ & 83.85 $\pm$ 0.44                      \\

	K (\ms) 	&506 $\pm$ 37	                           \\
    e 			& 0.0 (fixed)  	\\
    $\omega~(\deg)$ & 90 (fixed) \\
   Jitter (\ms)& 34$^{+39}_{-21}$                     \\
    \mpl (\mjup)& 2.36$\pm$0.21	\\
    \rpl (\rjup)& 1.33$\pm$0.03  \\
    $\rho_{p}$ (\gccc) & 1.25$\pm$0.15 \\
    a (AU) & 0.0236$\pm$0.0005 \\
    T$^{\ast}_{eq}$ (K) & 1357 $\pm$ 15
	\\
	\hline
    \multicolumn{2}{l}{$\ast$ Assumed zero Bond albedo}\\
	\end{tabular}
    \label{tab:planet}
\end{table}
\subsection{Stellar Rotation from NGTS data}
\label{sub:rotation}
\begin{figure}
	\includegraphics[width=\columnwidth,height=6cm,angle=0]{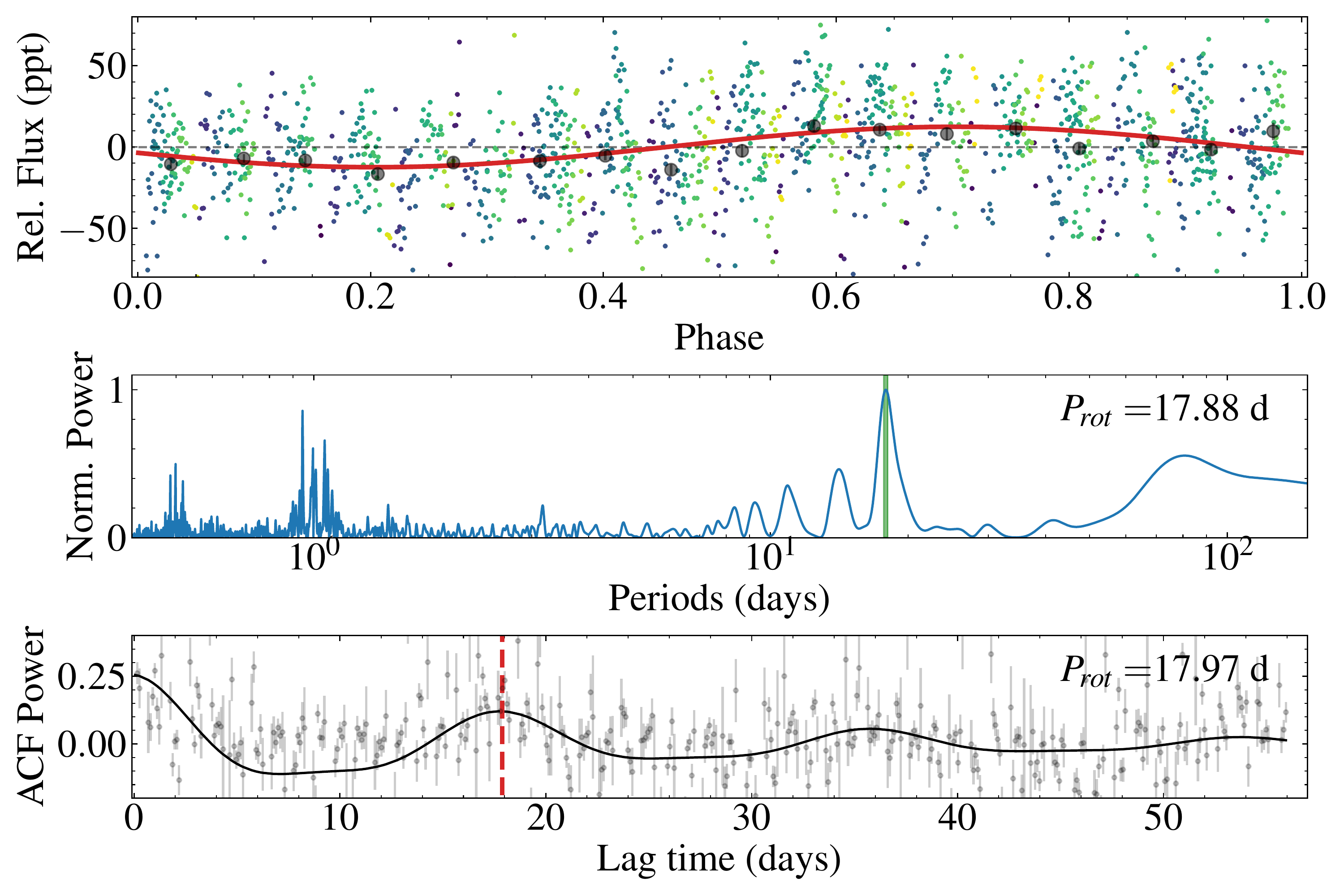}
    \caption{\textbf{Top}: NGTS lightcurve wrapped around the best-fitting LS period, where blue to green colors represent beginning to end of observations. Red line shows a sinusoidal model with amplitude $11.4^{+1.4}_{-1.6}$\,ppt. \textbf{Middle}: The Lomb-Scargle periodogram of the \Nplanet\ photometric data. The green bar centred at the highest peak displays the rotational period P$_{\rm rot} = 17.89 \pm 0.08$\,d.  \textbf{Bottom}: Lag time as a function of ACF power. Red vertical dashed line marks the 1$^\text{st}$ highest auto-correlation period at lag time 17.97 day, matching \Nstar\ rotation period.}
    \label{fig:rotation}
\end{figure}
The rotation period of stars can be measured by modelling the photometric brightness variation caused by starspots coming in and out of sight as stars spin. Thanks to several ground- and space-based missions, P$_\text{rot}$ measurements have been extracted for thousands of stars of distinct spectral types \citep{mcquillan2014rotation,martins2020search,briegal2022periodic}, thus helping set constraints on the dominant mechanisms driving stellar angular momentum evolution \citep{kawaler1988angular, bouvier2014angular}. Moreover, rotation periods are widely used to calibrate gyrochronology models \citep{angus2019toward}, which in turn, are used to infer stellar ages as a first order function of P$_\text{rot}$, yet limitations exist  \citep[][]{barnes2007ages,epstein2013good}. 

We extracted \Nstar\ rotation period with the Lomb-Scargle (LS) periodogram \citep{rebull2016rotation,vanderplas2018understanding} as well as auto-correlation functions (ACF) methods \citep{angus2018inferring}. Each technique has its own assumptions, advantages, and limitations, i.e, while the
LS method assumes a sinusoidal function to model the rotation signal, thus best suited for datasets presenting stable oscillations, the ACF technique is a more flexible method that measures the degree of similarity between different parts of the dataset \citep[see,][]{gillen2020ngts}.

Prior to the period search, we masked the transits and binned the data to 30 minute cadence. Both LS and ACF methods are applied to the dataset, which detect rotation periods whose difference is of $\delta \text{P}_{\rm rot}=0.1$\,d (Fig \ref{fig:rotation}). Since neither LS nor ACF techniques provide a confidence interval, a bootstrap approach was used to draw 15000 sample from the data with replacement. For each sample we fitted a sine model and compute the rotation period from the LS periodogram, thus generating distributions for both P$_{\text{rot}}$ and amplitude (A$_{\text{\rm rot}}$), where the median and 1$-\sigma$ intervals give P$_{\text{rot}} = 17.89 \pm 0.1$\,d and A$_{\text{rot}} = 10 \pm 1$\,ppt. In order to provide further confidence on our P$_{\rm rot}$ estimation, we attempted to measure the activity index ${\rm log_{10}R_{HK}}$ and projected rotational velocity ${\rm vsini}$, yet we were unable to estimate such parameters due to the spectrum low (< 10) SNR. Figure \ref{fig:rotation} (top) shows the phase-folded lightcurve to the period at maximum power of the LS periodogram (centre) from one bootstrap realisation with its corresponding sinusoidal model, while the rotation period from the ACF (bottom) was computed with the \texttt{astroML} package based on the Edelson $\&$ Krolik method \citep{edelson1988discrete}, where we adjusted an underdamped Simple Harmonic Oscillator (uSHO) to the data in order to extract the period. The LS method is available through the \texttt{astropy} package based on \citep{vanderplas2015periodograms}. \\
Finally, we visually checked the periods presenting moderately high LS power, and ruled out the ones below and near one day, which are likely associated to either instrumental noise or poor observing conditions. Yet, the $\sim$ 14.2 days signal near the rotation period cannot be associated to neither half the Lunar cycle of 14.8 days nor other non-astrophysical signals. Therefore, we followed the same procedure described above to model the 14.2 days signal, and found the best-fitting period and amplitude of 14.18 $\pm$ 0.13 d and 9.27 $\pm$ 0.98 ppt, respectively, and associate it to a possible \Nstar\ differential rotation. The stellar rotation period we derived was independently confirmed with the \texttt{RoTo}\footnote{https://github.com/joshbriegal/roto} code \citep{josh_briegal_2022_6994195}, which extracts stellar rotation periods automatically using LS, generalised ACF and GP, thus providing further confidence on our reported measurement.
\subsection{Transit Timing Variation analysis}
\label{sub:ttvmodeling}
\begin{figure}
	\includegraphics[width=\columnwidth]{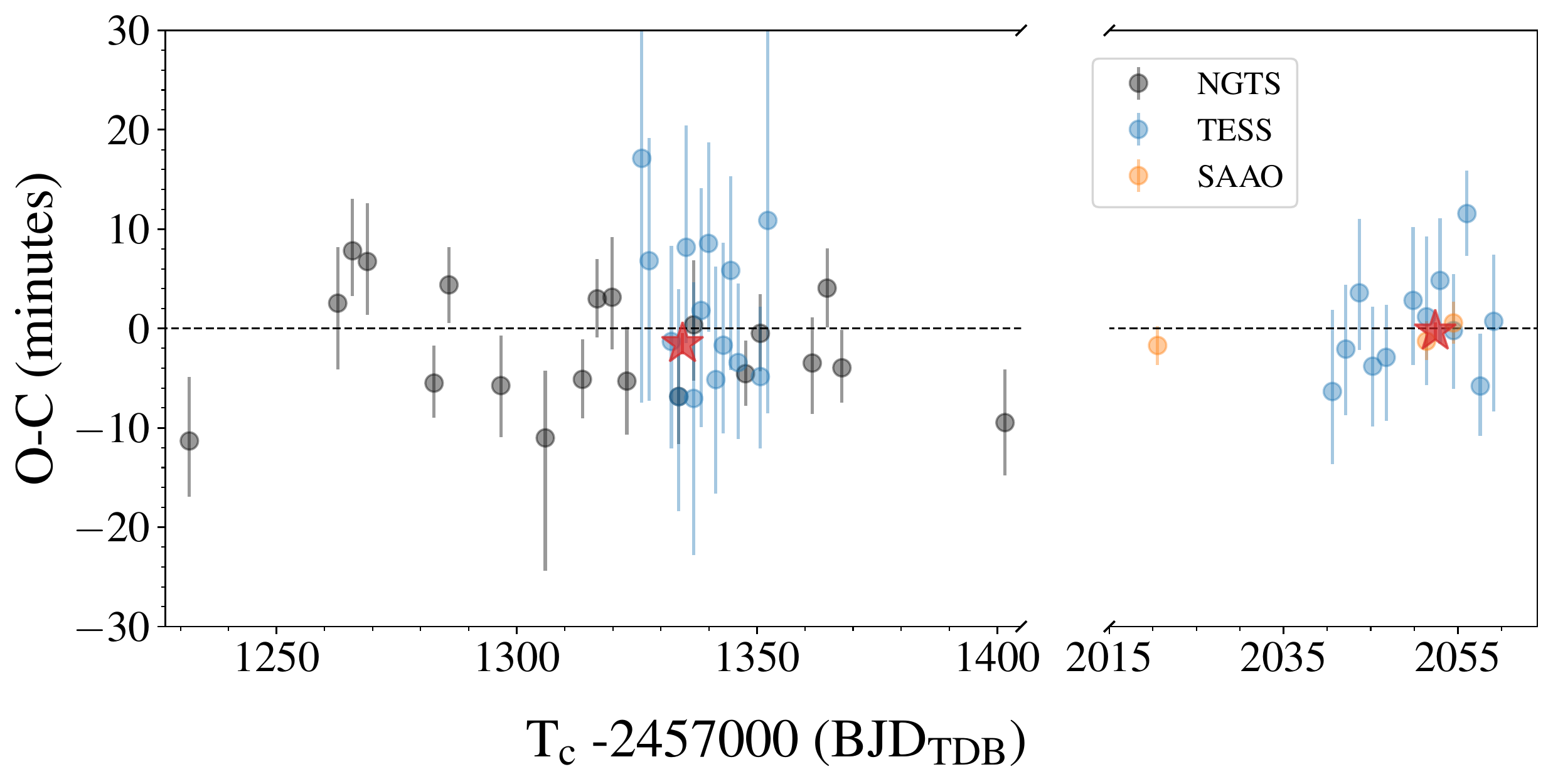}
    \caption{\Nplanet\ computed Transit Timing Variation. The zero dashed line indicates no deviation from the linear ephemeris model. Red stars represent the left and right portion of datasets average, with inverse variances as weights. 1. A table is available in a machine-readable format from the online journal.}
    \label{fig:ttvs}
\end{figure}
The Transit Timing Variation (TTV) method \citep{agol2005detecting} was responsible for the validation of several multi-planet systems around faint stars from the Kepler mission \citep{cochran2011kepler, gillon2017seven,steffen2012transit}. Since the majority of Kepler stars were faint, the RV method lacked enough precision to confirm the majority of the transits as bonafide planets, although a few had RV detections \citep[][]{barros2014sophie, almenara2018sophie}. Therefore, the TTV method became key to determine planetary masses/eccentricities for faint multi-planet systems \citep{lithwick2012extracting}. Moreover, extensive TTV/RV searches for hot Jupiter companions supported the hypothesis that such massive planets are not part of multi-planet system \citep{steffen2012kepler, holczer2016transit}, thus setting major constraints on giant planets orbital evolution.

The TTV method consists of measuring the difference between each mid transit T$_n$ from the expected transit time computed with a linear ephemeris model given by $T_n = T_0 + n \times P$, where $n$ and $P$ are the transit number and period, respectively. Deviation from the linear model are frequently associated to dynamical interactions, with the most common cases being planets near mean motion resonances \citep[e.g,][]{bryant2021transit} and planet-star tidal interaction leading to orbital decay \citep{yee2019orbit}.

\Nplanet\ observed transit times as well as the linear fit were done with \texttt{Juliet} while holding all parameters to the posterior median from table \ref{tab:planet} except for the set of $T_n$, which was given a normal prior $\mathcal{N}(T_{n}, 0.1^2)$. Figure \ref{fig:ttvs} shows the modelled observed transit times subtracted from the best-fitting linear model $T_n = (2458214.8890 \pm 0.0012) + n \times (1.5433891 \pm 0.0000033)$, which indicates an agreement between observed transits and the model.
%
\section{Discussion}
\label{sec:discussion}
\begin{figure}
	\includegraphics[width=\columnwidth,angle=0]{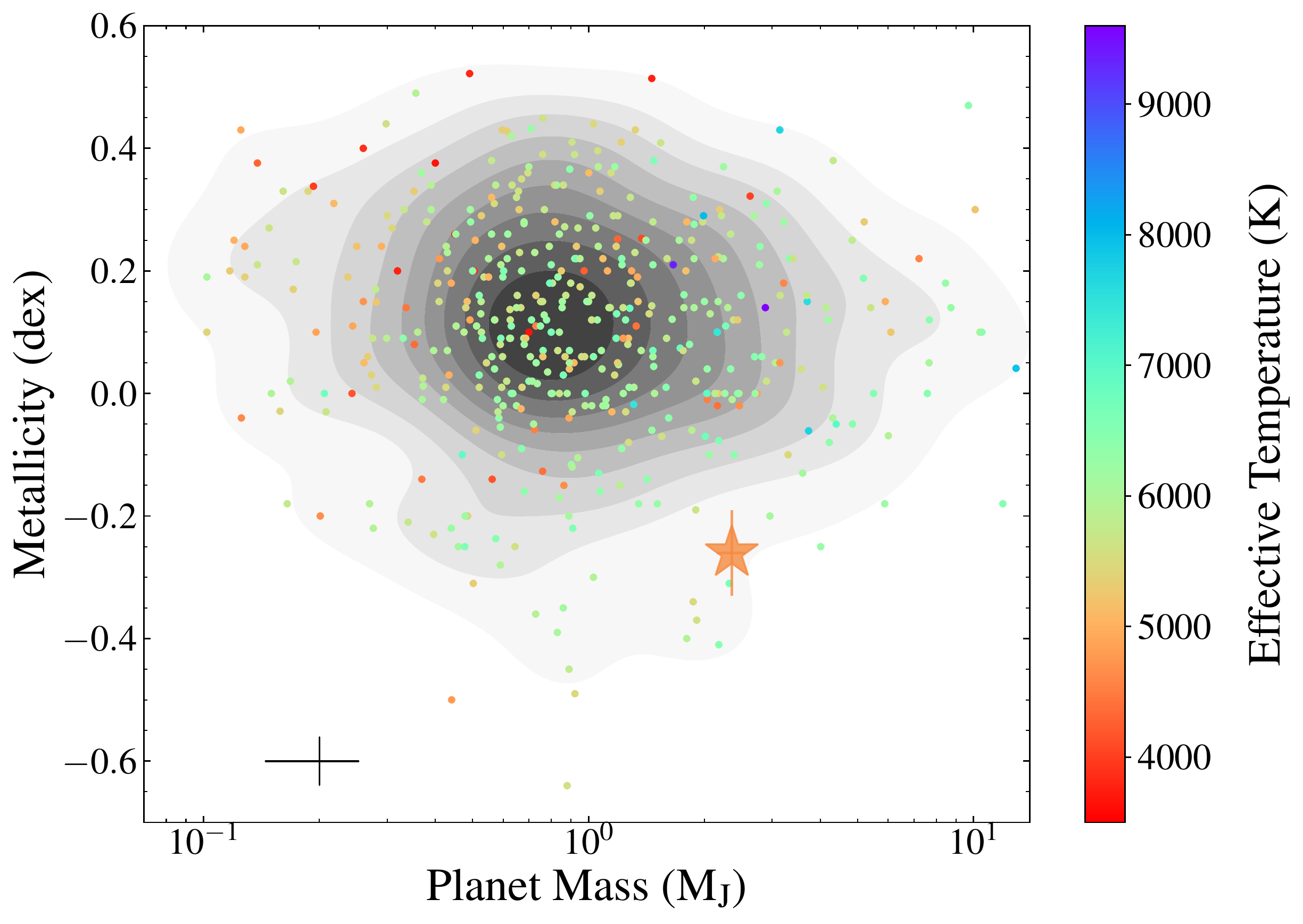}
    \caption{Transiting HJs planet mass against stellar metallicity. Stars are color-coded by their effective temperature, and \Nplanet\ is represented as an orange star symbol towards the bottom right of the plot. Dark to light shades of gray represent high to low planet number density, and black cross at the bottom left corner represents the standard deviation of planet mass and [Fe/H] uncertainties.}
    \label{fig:FeHMpMsPlot}
\end{figure}

Our data analysis (section \ref{sec:analysis}) reveals what is the first NGTS discovery of a rare planetary system composed of a massive planet hosted by a relatively metal-poor star. These properties place \Nstar\ at a heavily underpopulated region of the $\mjup$ vs [Fe/H] parameter space (Fig \ref{fig:FeHMpMsPlot}), and the only massive giant hosted by a K dwarf at that [Fe/H], thus making \Nstar\ a unique system.
\subsection{The stellar metallicity vs bulk density plane}
\label{subsub:fehrhoplane}
\begin{figure}
	\includegraphics[width=\columnwidth,angle=0]{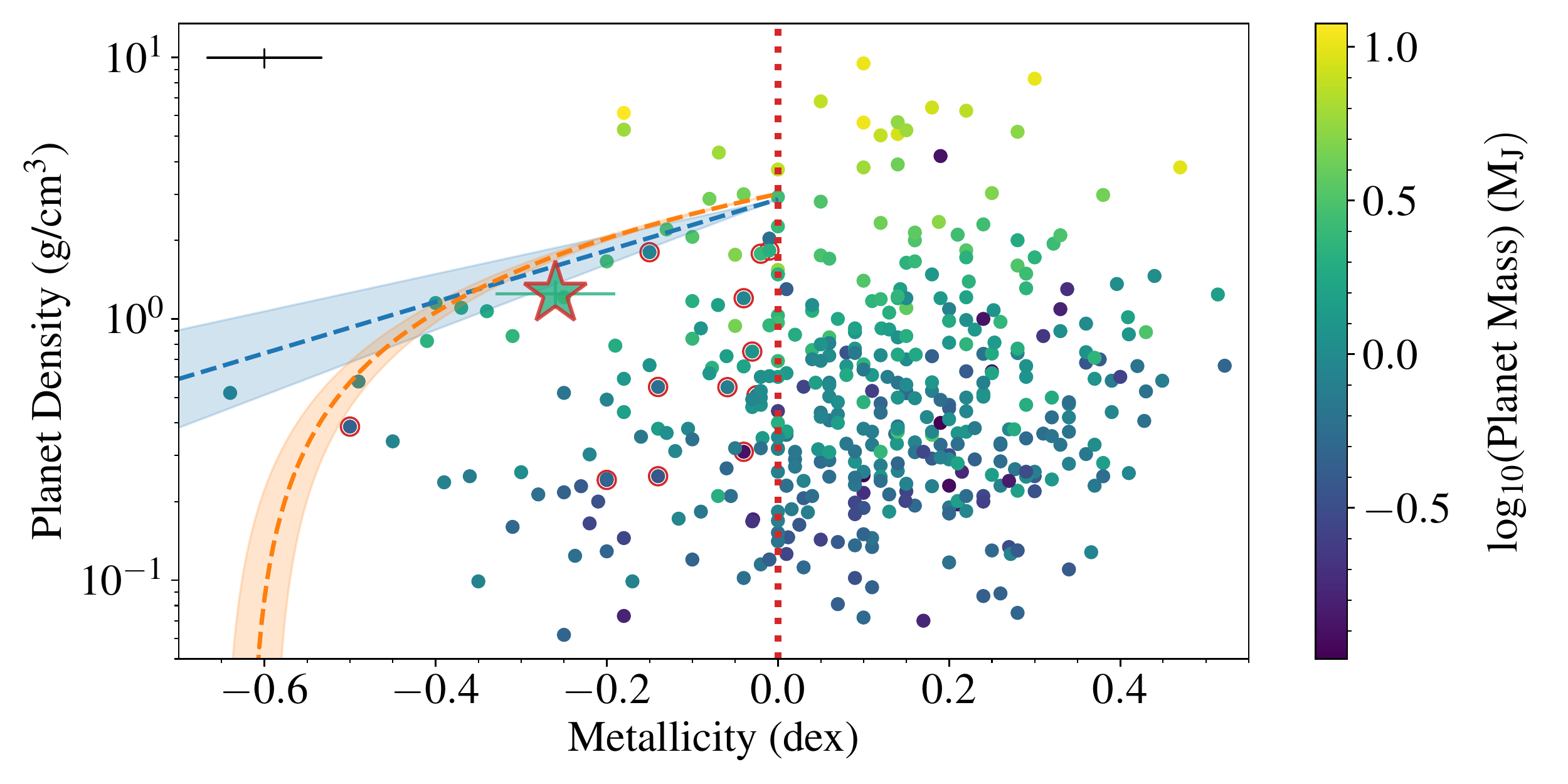}
    \caption{Stellar metallicity vs planet bulk density color-coded by the logarithm of planet mass. Red open circles show metal-poor ([Fe/H] < 0) K dwarf stars (T$_{\rm eff}$ = 3900-5200 K). The \Nstar\ system is indicated by the light green star at the upper left from centre, while the black cross at the top left corner represents the standard deviation of HJs uncertainties. Blue and orange dashed lines represent empirical exponential and linear models with shaded regions displaying their 1-$\sigma$ confidence intervals, respectively. The zero metallicity border is shown as red vertical dashed line}
    \label{fig:FeHVSrhop}
\end{figure}

Figure \ref{fig:FeHVSrhop} compares host star metallicity with planetary bulk density ($\rho_{\rm b}$) for a well-studied population of HJs from the TEPCat catalogue \citep[][]{southworth2011homogeneous}, all color-coded by planetary mass. Our analysis shows that \Nplanet\ is relatively dense ($1.25 \pm 0.15$\,g/cm$^3$) when compared to other HJs orbiting K dwarf metal-poor stars, and one of the densest amongst metal poor hosts below [Fe/H] < -0.2. Moreover, a clear upper boundary is observed, with planets $\rho_{\rm b}$ decreasing as an inverse function of stellar metallicity. Such a trend is in agreement with core-accretion models, whereby HJs formed in low metallicity environments would have smaller cores, and consequently lower bulk densities, which descends as even less metal content is present. Two empirical models were adjusted as an approximation to match this upper boundary reflecting the possible correlation between [Fe/H] and $\rho_{\rm b}$. The model in blue represents an exponential of the form $\rho_{\rm b} = a{\rm e}^{b\rm [Fe/H]}$, with $a$ and $b$ given by $2.9 \pm 0.1$ and $2.3 \pm 0.6$, while the linear model in orange was defined as $\rho_{\rm b} = c{\rm [Fe/H]} + d$, with $c$ and $d$ given by $4.9 \pm 0.2$ and $3.01 \pm 0.03$. Although we used empirical models to derive an upper boundary for HJs bulk densities, a larger sample of transiting HJs hosted by metal-poor stars, as well as a proper physical model to describe planet formation as a function of protoplanetary disc metallicity, are necessary to claim such correlation. Finally, a metal-poor gap may exist in the parameter space, with two classes of HJs orbiting metal-poor stars, the dominant population being low density and metal-poor HJs, and a less crowded population of higher density HJs, possibly large core-hosting planets. In order to confirm such a hypothesis, more transiting HJ planets are required within the metal-poor parameter space, particularly dense HJs, such that statistical samples can be drawn to test the reality of the gap.  
\subsection{Radius Inflation}
\label{subsub:radiusanomaly}
Several studies point to a high incident stellar flux as being the probable mechanism responsible for the HJs radius inflation, where energy is deposited into the planet interior, thus leading to an increase in radius. \citet{demory2011lack, miller2011heavy} shows that the physical mechanisms driving the radius anomaly operates above an incidence flux throughput of $\sim$ 2$\times10^5$ Wm${^{-2}}$ while \citet{thorngren2018bayesian} performed statistical analysis based on planetary thermal evolution models on a sample of 281 HJs, and showed the necessary conversion of incident flux to internal heating required to reproduce HJs observed radii peak at equilibrium temperature (T$_\text{eq}) \sim$ 1500 K. \citet{hartman2016hat} shows that HJ radii grows as a function of main sequence stars fractional ages, i.e, as stars age on the main sequence, they brighten up, thus leading to higher planetary irradiation and hence higher T$_\text{eq}$ of their orbiting planets. However, alternative scenario that could explain HJ radius anomalies such as star-planet tidal interactions, which lead to internal heating of the planet, thus causing a radius inflation \citep[e.g, see][for a review]{fortney2021hot}.

\begin{figure}
	\includegraphics[width=\columnwidth,angle=0]{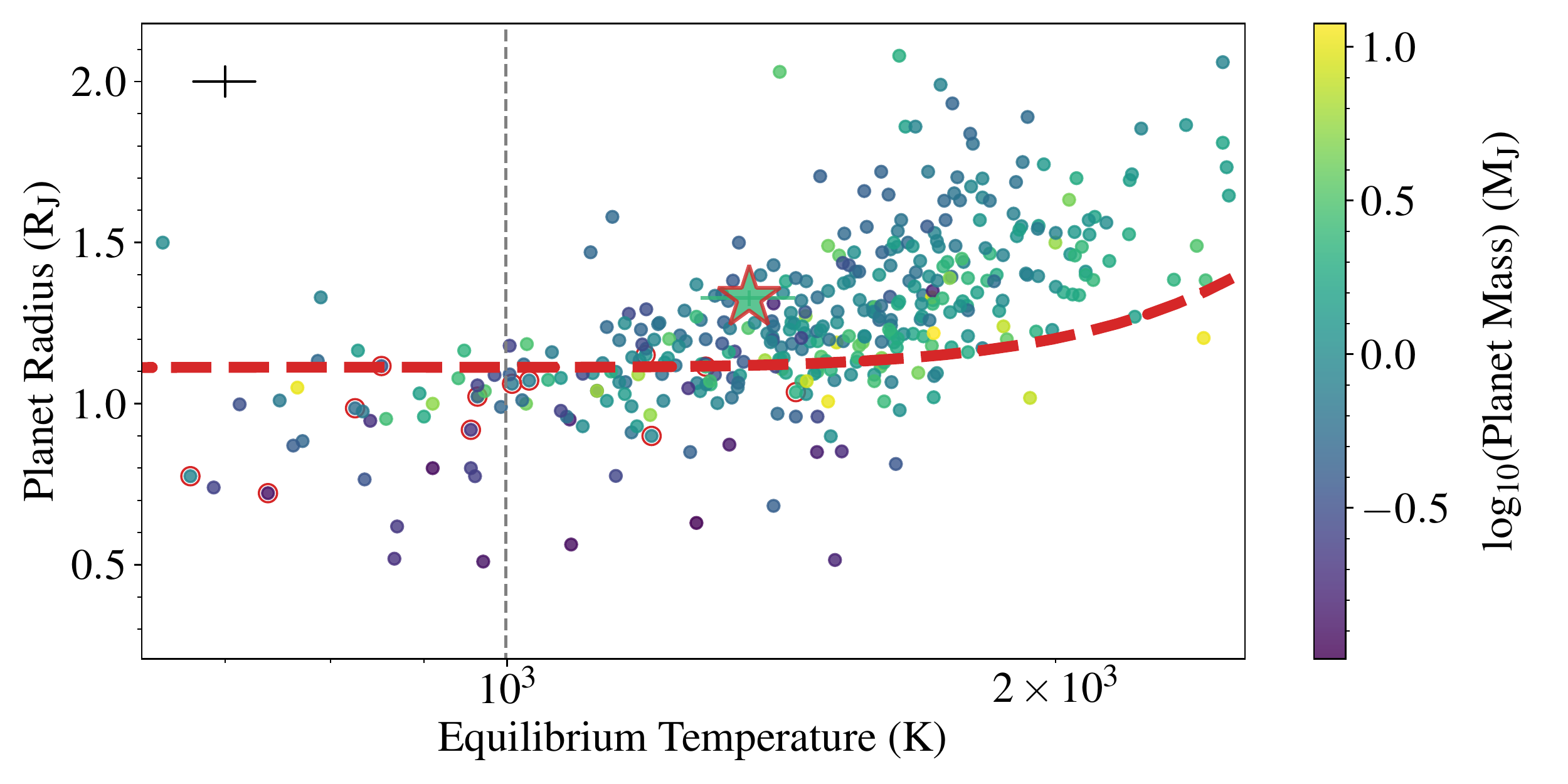}
	\includegraphics[width=\columnwidth,angle=0]{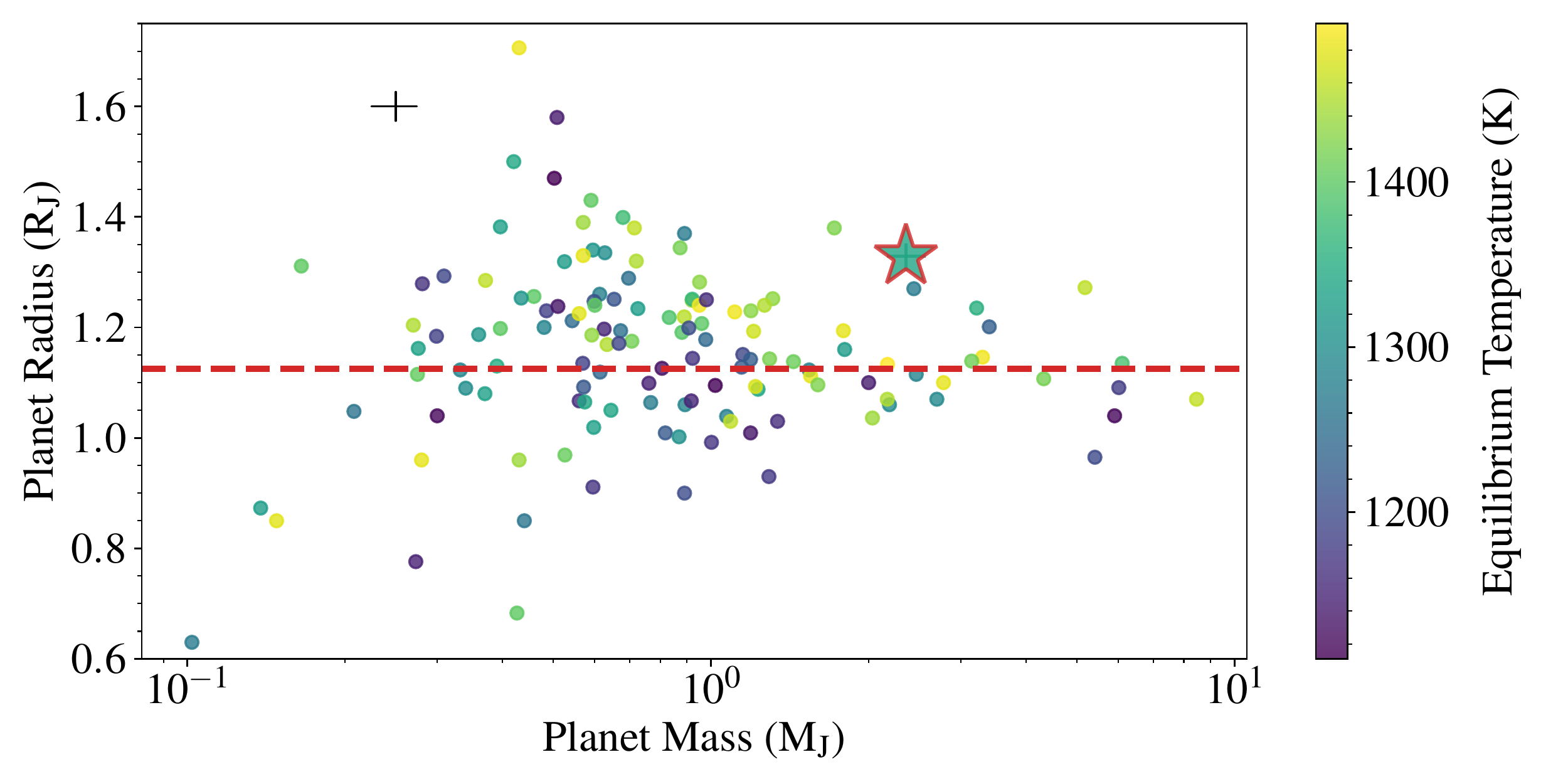}
    \caption{\textbf{Top}: Equilibrium temperature vs planet radius color-coded by the logarithm of planet mass. Red dashed line represents an inflation-free model for a HJ evolved to 4.5 Gyr with a H/He composition adapted from \citet{thorngren2018bayesian}. \Nplanet\ is displayed by the light green star above the model and near the image centre. Black cross at the top left corner represents the HJs parameters uncertainties standard deviation, and red open circles marks metal-poor K dwarf stars.
    \textbf{Bottom}: M$_{\rm p}$ vs R$_{\rm p}$ color-coded by T$_{\rm eq}$ showing HJs from the top figure with T$_{\rm eq}$ between 1100-1500 K. Red dashed line shows expected radius of 1.125 R$_{\rm J}$ from the same inflation-free model at \Nplanet\ T$_{\rm eq}$.}
    \label{fig:TeqVSRp}
\end{figure}

Figure \ref{fig:TeqVSRp} top compares giant planet equilibrium temperatures to their measured radii, where \Nplanet\ presents a rather large radius when compared to HJs with similar T$_{\rm eq} \sim 1300$\,K, and also when compared to an inflation-free model \citep{thorngren2018bayesian}, thus pointing to a possible inflated planet. The lower panel in the figure shows the lack of massive inflated HJs, which highlights the importance of confirming the inflated nature of \Nplanet. 

Planetary structure models by \citet{fortney2007planetary} (hereafter F07) predict \Nplanet\ to have a radius $\sim$ 21$\%$ smaller for an age of $\sim$ 3 Gyr, while \citet{baraffe2008structure} (hereafter B08) inflation-free models, which take into account the metal mass fraction (Z) and its distribution within the planetary interior, predicts a radius of $\sim$ 21$\%$ smaller at Z=0.02 and no planet irradiation. We also compared B08 models that consider stellar irradiation, thus giving an $\sim$ 16$\%$ smaller radius at 3 Gyr and Z=0.02. Neither F07 nor B08 predict a radius consistent with observations, unless the planetary system is very young (100-500 Myrs), where HJs radii are typically large, the models agree to our observed radius; yet we rejected the hypothesis that \Nstar\, is a young system (< 1 Gyr) in $\S$ \ref{sub:stellar}.

To further confirm the inflated nature of the planet, we followed the method described in \citet{costes2020ngts}, and based on \citet{sestovic2018investigating} (hereafter S18), where an empirical model relating the expected radius inflation $\Delta R$ to planet radius, mass and incident fluxes derived from Bayesian statistical analysis on a sample of 286 transiting HJs. Firstly, we estimated an incident flux (F) of $3.4 \pm 0.7 \times 10^{6}$ Wm${^{-2}}$ for \Nplanet\ from \citet{weiss2013mass} Eq. (9),
\begin{equation}
    \frac{\rpl}{R_{\oplus}} = 2.45\left(\frac{M_{\rm p}}{M_\oplus}\right)^{-0.039}\left(\frac{F}{\rm ergs~s^{-1}~cm^{-2}}\right)^{0.094}
\end{equation}
which is valid for M$_{\rm p} > 150$~M$_\oplus$, and from S18 Eq. (11),
\begin{equation}
    \Delta R = 0.52\left({\rm log_{10}}F - 5.8\right),~ 0.98 \leq \frac{M_{\rm p}}{M_{\rm J}} <2.5 
\end{equation}
we found a $\Delta R$ of $0.38 \pm 0.05$. The inflated radius is given by, $R_{\rm inf} = C + \Delta R$, where $C$ is the baseline radius from S18 Eq. (1), with best fit value of $1.06 \pm 0.03$ R$_{\rm J}$ from S18 Table 1. Therefore, a $R_{\rm inf}$ of $1.44 \pm 0.06$ R$_{\rm J}$ is expected for our estimated incidence flux on \Nplanet\,, and is in statistical agreement to our measured radius from Table \ref{tab:planet}.
\section{Conclusion}
\label{sec:concl}
We report the discovery of \Nplanet, a hot Jupiter with a mass, radius, and bulk density of $2.36 \pm 0.21$\,\mjup, $1.33 \pm 0.03$\,\rjup, and $1.25 \pm 0.15$\,g/cm$^3$, respectively. The planet orbits a K3V star every 1.5 days, representing one of the shortest period gas giants orbiting such a low-mass star, and its large mass also makes it one of the most massive HJs orbiting such a star. We also find the planet to be inflated by around 21\% when comparing to inflation-free planetary structure models, and is significantly larger than other similar gas giants with effective temperatures in agreement with that of \Nplanet. The close proximity of the planet to its host star means that a combination of stellar irradiation and tidal heating could explain the inflated nature of the planet's atmosphere.

When placing \Nplanet\ in the metallicity vs planet bulk density plane for HJs, we identify a falling upper boundary in the metal-poor regime.  The density of HJs decrease as a function of host star metallicity, which likely reflects the formation pathway for these planets. The large cores that are required to explain their high densities, drop in mass as a function of decreasing metallicity, since there exists less metals in the proto-planetary disc to quickly form larger cores through core accretion before the disc disperses. This decrease in core mass then returns a decrease in their bulk densities too.  We fit two empirical models to this upper envelope in order to better characterise the effect. We also find weak evidence for the existence of a gap in this part of the parameter space, yet more observations and better statistics are required to confirm the gap's existence.

The host star \Nstar\ shows moderately low activity, as evidenced by the light curves low spot modulation amplitudes, and absence of flare activity. Moreover, its age of $10.02^{+3.29}_{-7.30}$\, Gyr and rotation period $17.88 \pm 0.08$\,d are in accordance with expected ages of 1.0$\--$4.5 Gyr from gyrochronology models. A second rotation period was detected in the LS periodogram, thus indicating that \Nstar\ exhibits evidence for differential rotation. The planet’s transit times were extracted and fitted by a linear ephemeris model, with residuals showing no transit time variations. In addition, light curve eyeballing and BLS methods do not return any evidence of an additional companion in the system. 

The discovery of \Nplanet\ will add to the small yet increasing population of massive HJ planets around low-mass and metal-poor stars, thus helping place further constraints on current formation and evolution model for such planetary systems.

\section*{Acknowledgements}
Based on data collected under the NGTS project at the ESO La Silla Paranal Observatory. The NGTS facility is operated by the consortium institutes with support from the UK Science and Technology Facilities Council (STFC) under projects ST/M001962/1, ST/S002642/1 and ST/W003163/1. 
This study is based on observations collected at the European Southern Observatory under ESO programme 105.20G9.
DRA acknowledges support of ANID-PFCHA/Doctorado Nacional-21200343, Chile. 
JSJ greatfully acknowledges support by FONDECYT grant 1201371 and from the ANID BASAL projects ACE210002 and FB210003.
JIV acknowledges support of CONICYT-PFCHA/Doctorado Nacional-21191829.
Contributions at the University of Geneva by ML, FB and SU were carried out within the framework of the National Centre for Competence in Research "PlanetS" supported by the Swiss National Science Foundation (SNSF).
The contributions at the University of Warwick by PJW, SG, DB and RGW have been supported by STFC through consolidated grants ST/P000495/1 and ST/T000406/1.
The contributions at the University of Leicester by MGW and MRB have been supported by STFC through consolidated grant ST/N000757/1.

CAW acknowledges support from the STFC grant ST/P000312/1.
TL was also supported by STFC studentship 1226157.
MNG acknowledges support from the European Space Agency (ESA) as an ESA Research Fellow.
This project has received funding from the European Research Council (ERC) under the European Union's Horizon 2020 research and innovation programme (grant agreement No 681601).
The research leading to these results has received funding from the European Research Council under the European Union's Seventh Framework Programme (FP/2007-2013) / ERC Grant Agreement n. 320964 (WDTracer).
The contribution of ML has been carried out within the framework of the NCCR PlanetS supported by the Swiss National Science Foundation under grants 51NF40\_182901 and 51NF40\_205606. ML also acknowledges support of the Swiss National Science Foundation under grant number PCEFP2\_194576.

\section*{DATA AVAILABILITY}

The data underlying this article are made available in its online supplementary material.




\bibliographystyle{mnras}
\bibliography{paper} 








\bsp	
\label{lastpage}
\end{document}